\documentclass[pre,aps,reprint,showpacs]{revtex4-1}

\usepackage{graphicx}
\usepackage{amsmath}
\usepackage{amssymb}
\usepackage{mathtools}
\usepackage[american]{babel}
\usepackage{hyperref}

\newcommand{\nn}{\nonumber\\}
\newcommand{\mean}[1]{\left\langle #1 \right\rangle}
\newcommand{\sgn}{\text{sign}}
\newcommand{\expO}[1]{\exp\left[#1\right]}

\newcommand{\peq}[1]{P_{#1}}
\newcommand{\domain}{\mathcal{D}}
\newcommand{\ot}[2]{{#2}_{#1}}
\newcommand{\otf}[2]{\Delta {#2}_{#1}}
\newcommand{\Ot}[2]{T_{#1}\left(#2\right)}
\newcommand{\Otf}[2]{\Delta T_{#1}\left(#2\right)}
\newcommand{\infd}{\mathcal{I}}
\newcommand{\infdens}[1]{\infd\left(#1\right)}
\newcommand{\infdensg}[2]{\infd_{#2}^>\!\left(#1\right)}
\newcommand{\infdensl}[2]{\infd_{#2}^<\!\left(#1\right)}
\newcommand{\infdensgl}[2]{\infd_{#2}^\gtrless\!\left(#1\right)}
\newcommand{\subbulk}{\mathrm{B}}
\newcommand{\subldev}{\mathrm{L}}
\newcommand{\subnx}{{\text{n}x}}
\newcommand{\suprw}{{\text{RW}}}
\newcommand{\kb}{k_\text{B}}
\newcommand{\temp}{\mathcal{T}}
\newcommand{\pot}[1]{V_{#1}}
\newcommand{\Order}[1]{O\left[#1\right]}
\newcommand{\trafo}[1]{\widehat{#1}}
\newcommand{\trafoF}[1]{\widetilde{#1}}
\newcommand{\mql}{M_q^<}
\newcommand{\mqg}{M_q^>}
\newcommand{\Obs}{\mathcal{O}}
\newcommand{\obs}{o}
\newcommand{\Obsf}{\Delta\Obs}
\newcommand{\obsf}{\Delta\obs}
\newcommand{\ta}[1]{\overline{#1}}
\newcommand{\indfu}[1]{\mathbf{1}\!\left\{#1\right\}}
\newcommand{\meaneq}[1]{\left\langle{#1}\right\rangle_{\!\mathrm{eq}}}

\newcommand{\meaneqcond}[2]{\left\langle{#1}\right\rangle_{\!\mathrm{eq},\;#2}}

\begin{document}

\title{Fluctuations around equilibrium laws in ergodic continuous-time random 
walks}

\author{Johannes H. P. Schulz}
\affiliation{Department of Physics, Institute of Nanotechnology and Advanced 
Materials, Bar-Ilan University, Ramat Gan 52900, Israel}
\author{Eli Barkai}
\affiliation{Department of Physics, Institute of Nanotechnology and Advanced 
Materials, Bar-Ilan University, Ramat Gan 52900, Israel}

\date{\today}

\begin{abstract}
We study occupation time statistics in ergodic continuous-time random walks. 
Under thermal detailed balance conditions, the average occupation time is given 
by the Boltzmann-Gibbs canonical law. But close to the non-ergodic phase, the 
finite-time fluctuations around this mean are large and nontrivial. They exhibit 
dual time scaling and distribution laws: the infinite density of large 
fluctuations complements the L\'evy-stable density of bulk fluctuations. Neither 
of the two should be interpreted as a stand-alone limiting law, as each has its 
own deficiency: the infinite density has an infinite norm (despite particle 
conservation), while the stable distribution has an infinite variance (although 
occupation times are bounded). These unphysical divergences are remedied by 
consistent use and interpretation of both formulas. Interestingly, while the 
system's canonical equilibrium laws naturally determine the mean occupation time 
of the ergodic motion, they also control the infinite and L\'evy-stable 
densities of fluctuations. The duality of stable and infinite densities is in 
fact ubiquitous for these dynamics, as it concerns the time averages of general 
physical observables.
\end{abstract}



\maketitle

\section{Introduction}\label{sec.introduction}
Stochastic theories of motion are a well-established approach to model physical 
trajectories of single particles embedded in a thermal environment. The ergodic 
motions in particular are an essential prerequisite of statistical mechanics, as 
they exhibit the following convergence. Let, for a given trajectory, the 
occupation time $T_\domain(t)$ designate the total amount of time the particle 
has spent within a domain $\domain$ up to time $t$. Then
\begin{equation}\label{ergodic}
  \lim_{t\rightarrow\infty} \frac{\Ot{\domain}{t}}{t} = \peq{\domain} .
\end{equation}
$\peq{\domain}$ is the steady-state probability for the occupation of $\domain$. 
For example, for the overdamped motion of a Brownian particle immersed in water 
at temperature $\temp$ and subject to a static, confining potential $\pot{x}$, 
the Boltzmann-Gibbs canonical law yields $\peq{\domain}\propto\sum_{x\in 
\domain}\expO{-\pot{x}/(\kb\temp)}$; $\kb$ is Boltzmann's constant.

In ensemble theories, the focus rests on ensemble laws such as $\peq{\domain}$. 
But in practice, the measurement time $t$ is finite, so it is natural to 
investigate also the fluctuations
\begin{equation}
 \Otf{\domain}{t} = \Ot{\domain}{t} - tP_\domain .
 \label{otf}
\end{equation}
Knowing their precise distribution is relevant, e.g., when tracking single 
particles by confocal microscopy or when studying reaction-diffusion 
dynamics~\cite{Agmon2010,Berezhkovskii1998,Benichou2003}. A submanifold of the 
complete phase space is considered in the problem of phase 
persistence~\cite{Godreche2001,Majumdar1999}. The time a laser-cooled atom 
resides in the ``dark'' low-momentum state~\cite{Bardou2002,Bardou1994} 
determines the cooling efficiency. In a broader context, fluctuation 
theorems~\cite{Jarzynski2011,Seifert2012} describe deviations from equilibrium 
and refine our understanding of thermodynamic laws. As a rule of thumb, 
occupation time fluctuations are particularly large in the presence of 
annealed~\cite{Monthus1996,Bel2005,Bel2006} or 
quenched~\cite{Majumdar2002,Burov2007} disorder. The essential questions in all 
cases are: how large are the fluctuations, how do they evolve with time, and how 
are they determined by the precise microscopic dynamics on the one hand and the 
universal equilibrium laws $\peq{\domain}$ on the other?

We study here the site occupation times $\Ot{x}{t}$ of a confined 
continuous-time random walk (CTRW) on a lattice, $x\in\mathbb{Z}$. CTRWs and 
related models are a standard theoretical approach to describe dynamics where 
trapping mechanisms induce a variety of remarkable motion patterns; examples are 
the charge carrier transport in amorphous semiconductors~\cite{Scher1975}, model 
glasses~\cite{Monthus1996}, subrecoil laser 
cooling~\cite{Bardou2002,Bardou1994}, atomic transport in optical 
lattices~\cite{Lutz2004} and diffusion in biological 
cells~\cite{Jeon2011,Weigel2011,Wong2004,Xu2011}, to name a 
few~\cite{Bertin2008,Bouchaud1990,Metzler2000,Metzler2004,Metzler2014,RaWRaE}. 
We focus on a regime where the ergodic convergence~\eqref{ergodic} holds and the 
classical Boltzmann-Gibbs equilibrium and ergodicity ideas apply. Remarkably, 
the equilibrium probability $\peq{x}$ even controls the distribution of 
finite-time fluctuations~\eqref{otf}. 

But close to the non-ergodic regime, these fluctuations can be unexpectedly 
large, and their distribution exhibits a non-trivial, dual behavior. Firstly, 
the L\'evy-Gauss central limit theorem suggests a standard scaling approach, 
which yields the non-Gaussian L\'evy-stable laws. Secondly, it turns out that we 
have to explicitly consider the large fluctuations, which deviate from the 
central limit theorem. To study this aspect, we propose a nonstandard scaling 
procedure. It leads to a concept which may appear surprising in a probabilistic 
model: the infinite density. This density is not normalizable, despite being a 
limit of a properly normalized probability law. While such objects play a key 
role in the mathematical field of infinite ergodic 
theory~\cite{Thaler1983,Thaler2006,InfErgTheo}, their usage is less common in 
physical models, such as sub{\-}diffusion on intermittent 
maps~\cite{Korabel2009,Korabel2012,Akimoto2010,Akimoto2012}, diffusion in 
logarithmic potentials~\cite{Kessler2010,Holz2013,Lutz2013} and recently strong 
anomalous diffusion~\cite{Rebenshtok2014}. 
Our goal is to showcase the scope and applicability of infinite densities and 
familiarize the reader with its peculiar properties. We integrate it in the 
statistical mechanics framework by relating all our findings to the system's 
equilibrium statistics $\peq{x}$. Finally, we extend our discussion to include 
the statistics of time-averaged observables. This exposes the ubiquity of the 
interplay between dual scaling, stable laws and infinite densities in ergodic 
CTRWs.

We start by introducing the CTRW in Sec.~\ref{sec.model} and briefly review the 
concept of equilibrium in the CTRW context. The basic nature of the time scaling 
duality is explained in Sec.~\ref{sec.duality}. After a general introduction to 
the calculation of occupation time statistics and their long-time asymptotic 
approximation in Sec.~\ref{sec.otf}, we discuss in detail the emergent stable 
law of bulk fluctuations in Sec.~\ref{sec.otf:bulk}, and the infinite density of 
large fluctuations in Sec.~\ref{sec.otf:ldev}. Section~\ref{sec.moments} tackles 
the question of how to apply these dual limiting laws when it comes to ensemble 
averaging. In Sec.~\ref{sec.taobs}, we consider fluctuations of a time average 
of an arbitrary observable. To conclude, we summarize our findings and discuss 
briefly further directions of research, in particular models and applications 
beyond CTRW, in Sec.~\ref{sec.outlook}.

\section{The CTRW model and thermal equilibrium}\label{sec.model}
We consider a CTRW evolving along a one-dimensional lattice. To each accessible 
lattice point $x\in\mathbb{Z}$ we assign a probability $0< q_x< 1$ to make a 
jump to the right neighboring site; $1-q_x$ for jumps to the left. The jumps are 
instantaneous, but in between jumps, the particle resides on the lattice site 
for a random trapping or waiting time.  All jumps and waiting times are assumed 
to be statistically independent. The key quantity here is the common probability 
density function (PDF) $\psi(\tau)$ of waiting times $\tau$, which reflects the 
disorder and heterogeneity of the medium.
According to CTRW theory~\cite{RaWRaE}, the long-time qualities of the motion 
are related to the width of $\psi(\tau)$; more precisely, to the finiteness of 
its moments $\mean{\tau^n}=\int_0^\infty \tau^n\psi(\tau)\,d\tau$. Motivated by 
the complex systems collected in the introduction, we consider here 
distributions with power law tails, i.e. for large $\tau$
\begin{equation}
  \psi(\tau)\sim \frac{A}{|\Gamma(-\alpha)|\,\tau^{1+\alpha}} ,
  \label{broadtail}
\end{equation}
with the coefficient $A>0$, bearing the unit $\text{sec}^\alpha$. Some 
experimental setups permit the fine-tuning of the tail exponent $\alpha$ via 
physical parameters such as temperature~\cite{Xu2011} or particle 
size~\cite{Wong2004}. If the tail exponent is from the range $0<\alpha<1$, the 
mean waiting time $\mean{\tau}$ diverges. This regime has been investigated 
thoroughly in the past, as it gives rise to a wealth of anomalous 
phenomena~\cite{Monthus1996,Schulz2013b,Rebenshtok2008,Burov2010a}. We focus on 
$1<\alpha<2$, where waiting times do possess a characteristic relaxation scale, 
namely the mean $\mean{\tau}<\infty$. Consequently, as shown below, a confined 
CTRW is ergodic in the sense of Eq.~\eqref{ergodic}.
\begin{figure}
 \includegraphics[]{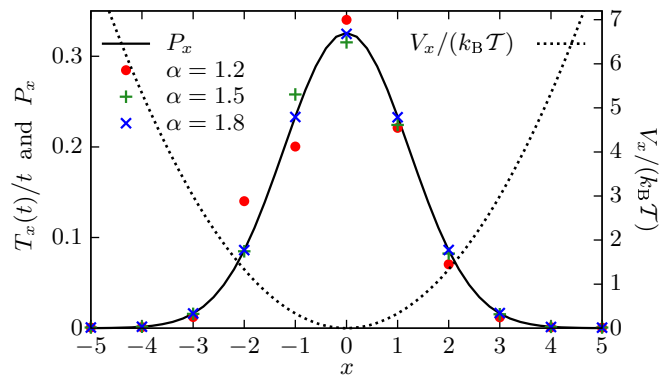}
 \caption{Occupation time fractions $\Ot{x}{t}/t$ of a bounded, thermal CTRW in 
comparison with ensemble occupation probabilities $\peq{x}$. The waiting time 
distribution is as in Eq.~\eqref{broadtail}, with $A=0.5$ and $\mean{\tau}=1$ 
(arbitrary units). The statistics of three different paths are shown (symbols), 
each for one value of $1<\alpha<2$ (see key). The $\peq{x}$ are Boltzmann's 
equilibrium law, Eq.~\eqref{boltzmann}, for the confining harmonic potential 
$\pot{x}/(\kb\temp)=x^2/3$. The occupation time fractions $\Ot{x}{t}/t$ roughly 
fall onto ensemble laws $\peq{x}$. But especially for the smaller $\alpha$, the 
deviations are huge, considering the long simulation time $t=10^6$.}
 \label{fig.average}
\end{figure}
In this context, $\peq{x}$ is the probability to find one random walker, picked 
from a large ensemble of non-interacting random walkers, on position $x$ after a 
large relaxation time $t/\mean{\tau}\rightarrow\infty$. But while the 
statement~\eqref{ergodic} is mathematically concise and correct, it may conceal 
physical reality. Figure~\ref{fig.average} compares occupation time fractions 
$\Ot{x}{t}/t$ with ensemble probabilities $\peq{x}$ for a CTRW particle which is 
bound by a harmonic potential $V_x\propto x^2$ and coupled to a heat bath at 
temperature $\temp$. In this case, the Boltzmann-Gibbs canonical ensemble laws 
imply a Gaussian form of $\peq{x}$ --- we discuss thermal systems in more detail
below. In Fig.~\ref{fig.average}, we observe a basic agreement with 
Eq.~\eqref{ergodic} on average, but deviations are significant at the huge, yet 
finite process time $t/\mean{\tau}=10^6$. The cause for the scatter are the 
waiting time statistics~\eqref{broadtail}, which, with $1<\alpha<2$, are so 
broad that $\mean{\tau^2}=\infty$. We note that Kac's theorem~\cite{Kac} in 
chaos theory~\cite{Zaslavsky2002} suggests that Hamiltonian systems should be 
studied under the premise $\mean{\tau}<\infty$, while $\mean{\tau^2}$ might 
indeed be infinite. Hence, we conduct below a detailed analysis of the 
occupation time fluctuations~\eqref{otf}.

\begin{figure*}
 \includegraphics[]{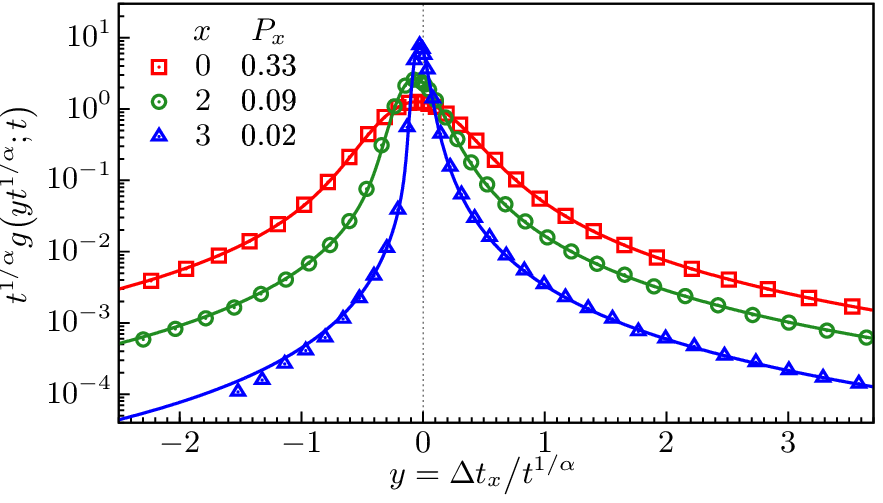}\includegraphics[]{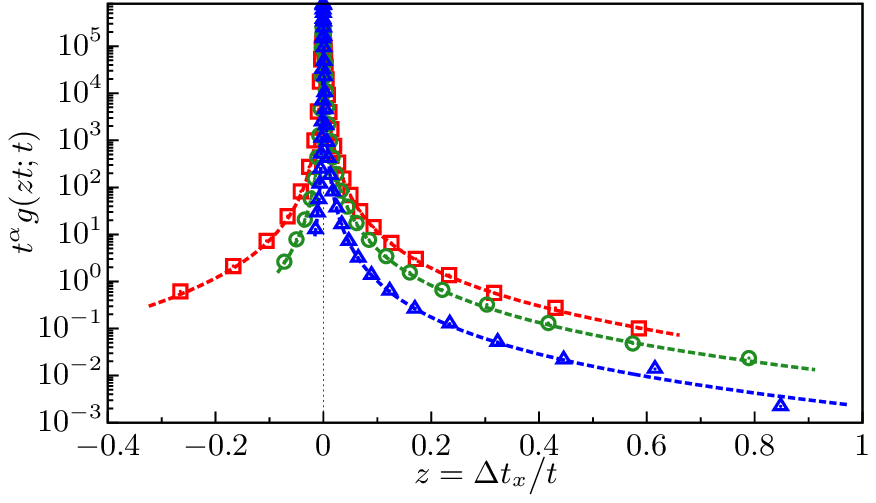}
 \caption{Scaled PDF of occupation time fluctuations $\Otf{x}{t}$ for confined 
CTRW. Symbols in both panels are the simulation results from $10^7$ trajectories 
with $\alpha=1.5$, evaluated on three lattice sites $x\in\{0,2,3\}$ (see key). 
All other parameters are as in Fig.~\ref{fig.average}. Both panels depict the 
same data. But depending on the time scaling we use, different forms of 
asymptotic behavior become apparent. \textit{Left:} Scaling with $t^{1/\alpha}$ 
as in Eq.~\eqref{otf:pdf:bulk:scalinglimit} yields the asymmetric, stable bulk 
statistics $g_\subbulk(\cdot)$ of Eq.~\eqref{otf:pdf:bulk:asymptotic} 
(continuous lines). The agreement between the asymptotic theory and the data is 
excellent. \textit{Right:} The nonstandard time scaling in 
Eq.~\eqref{otf:pdf:ldev:scalinglimit} produces the infinite density 
$g_\subldev(\cdot)$ for the large fluctuations as in 
Eq.~\eqref{otf:pdf:ldev:asymptotic} (dashed lines). These asymptotics are not 
normalizable due to the blow-up at the origin.}
 \label{fig.otf.pdf}
\end{figure*}

Under confinement, the probability to find the random walker at a site $x$ 
converges with time $t$ to the $t$-independent value $\peq{x}$. In this 
stationary state, the detailed balance condition holds,
\begin{equation}
  q_x\peq{x}=(1-q_{x+1})\peq{x+1} ,\qquad x\in\mathbb{Z} .
  \label{detbal}
\end{equation}
The existence of a stationary limiting state and detailed balance conditions are 
a premise for all CTRW dynamics studied in this work.
We may, in particular, choose to consider thermal CTRWs. Then, the hopping 
probabilities satisfy~\cite{Metzler2004,Bel2006} 
$q_x/(1-q_{x+1})=\expO{-(\pot{x+1}-\pot{x})/(\kb\temp)}$, $x\in\mathbb{Z}$. The 
aim of a thermal model is to describe systems where the particle is confined by 
an external potential $V_x$ and in contact with a heat bath at constant 
temperature $\temp$. Detailed balance then implies Boltzmann's canonical 
distribution
\begin{equation}\label{boltzmann}
 P_x=\frac{1}{Z}\exp\left(-\frac{\pot{x}}{\kb\temp}\right), \quad Z=\sum_x 
\exp\left(-\frac{\pot{x}}{\kb\temp}\right).
\end{equation}

Equations~\eqref{detbal}-\eqref{boltzmann} define a mapping between the sets 
$\{q_x\}$, $\{\peq{x}\}$ and $\{\pot{x}\}$, thus relating microscopic CTRW 
dynamics to the Boltzmann-Gibbs equilibrium concept. Recipes for the 
implementation in numerical simulations and a careful account on the effects of 
the lattice discretization can be found in Ref.~\cite{Bel2006}, Sec.~VIa.

\section{The dual time scaling of occupation times}\label{sec.duality}
Let  $\Ot{x}{t}$ be the occupation time at fixed  $x$ up to time $t$ since the 
start of the random walk at time $0$. We denote with $g(\otf{x}{t};t)$ the PDF 
of fluctuations $\Otf{x}{t}$ as in Eq.~\eqref{otf}. Its exact form is 
complicated and depends in particular on initial conditions, the full waiting 
time PDF $\psi(\tau)$ and the hopping probabilities $q_x$. But close to 
equilibrium, we expect to find a scaling limit form in terms of long-time 
properties of the process, such as the tail~\eqref{broadtail} and the 
equilibrium statistics $\peq{x}$. Surprisingly, the occupation times of CTRW 
feature not one, but two such asymptotic scaling forms. We derive below a bulk 
(``B'') scaling limit 
\begin{align}
  \label{otf:pdf:bulk:scalinglimit}
  g(\otf{x}{t};t) 
  &\sim 
\frac{1}{t^{1/\alpha}}\,g_\subbulk\!\left(\frac{\otf{x}{t}}{t^{1/\alpha}}\right) 
,\\
\intertext{while the large (``L'') fluctuations scaling limit is}
  \label{otf:pdf:ldev:scalinglimit}
  g(\otf{x}{t};t) 
  &\sim \frac{1}{t^{\alpha}}g_\subldev\!\left(\frac{\otf{x}{t}}{t}\right)
  .
\end{align}
Figure~\ref{fig.otf.pdf} demonstrates that these scaling laws do indeed coexist. 
The scaling functions $g_\subbulk(\cdot)$ and $g_\subldev(\cdot)$ are derived 
and discussed in Secs.~\ref{sec.otf:bulk} and~\ref{sec.otf:ldev}, respectively. 
But before going deeply into the analytical study, we want to develop an 
understanding for the character of this duality.

The total occupation time is a sum of independent waiting times with the broad 
statistics~\eqref{broadtail}. For the bulk, that is, the large majority of 
process realizations, the number of waiting times entering this sum is large, as 
its average is $t/\mean{\tau}$. By virtue of the generalized central limit 
theorem~\cite{GCLT}, one can thus indeed anticipate a scaling limit of the form 
of Eq.~\eqref{otf:pdf:bulk:scalinglimit}. We note that the bulk 
relation~\eqref{otf:pdf:bulk:scalinglimit} also has a common shorthand notation: 
$\Otf{x}{t}\sim t^{1/\alpha}$. Relations of this form appear in abundance in the 
physical and other sciences (with various $g_\subbulk(\cdot)$ and $\alpha$). It 
implies a strong statement on the asymptotic statistics: let a threshold 
$c(t)>0$ increase faster than $t^{1/\alpha}$; then the probability to observe 
$|\Otf{x}{t}|>c(t)$ goes to zero with $t$. In this sense, the scope of the 
approximation $g_\subbulk(\cdot)$ widens, and deviations from bulk behavior 
become rare. Hence, in many other physical contexts, a scaling relation of this 
type fully describes all asymptotic aspects of the respective process.

But for the CTRW problem at hand, the bulk analysis misses essential features of 
the distribution. The broad waiting time PDF~\eqref{broadtail} implies that the 
rare, large occupation times are relevant events. Since we can in principle have 
$0\leq\Ot{x}{t}\leq t$, the largest occupation times can be of the order of the 
measurement time $t$ itself. This is why we divide $\otf{x}{t}$ by $t$ in the 
large fluctuations scaling limit, Eq.~\eqref{otf:pdf:ldev:scalinglimit}. But 
notice that the latter has  a nonstandard, counter-intuitive form. The involved 
scaling function $g_\subldev(\cdot)$ is not properly normalized: integrating 
Eq.~\eqref{otf:pdf:ldev:scalinglimit} over all $\otf{x}{t}$ gives $1$ on the 
left-hand side, but on the right-hand side we get the seemingly time-dependent 
value $\int t^{-\alpha} g_\subldev(\otf{x}{t}/t)\,d(\otf{x}{t})\equiv 
t^{1-\alpha}\int g_\subldev(z)\,dz$. In fact, we will find in 
Sec.~\ref{sec.otf:ldev} that $g_\subldev(\cdot)$ has an infinite norm. Hence, it 
cannot be treated as a conventional PDF and we call it an infinite density. It 
is a useful and meaningful object, and contains exactly the statistical 
information that escapes the bulk scaling analysis.

\begin{figure}
 \includegraphics{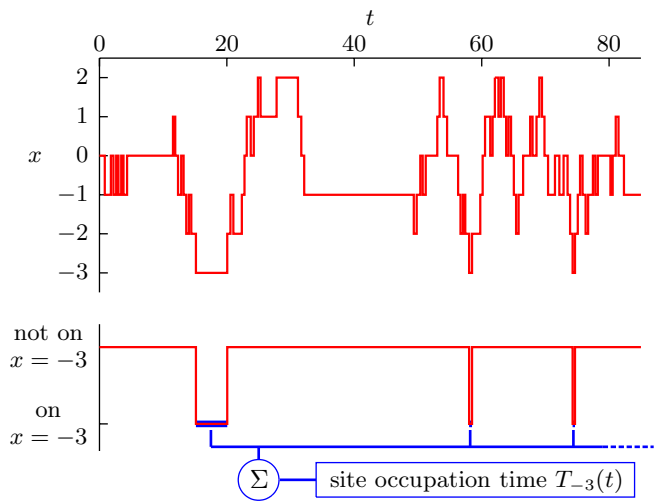}
 \caption{\textit{Top:} Trajectory of a CTRW with $\alpha=1.5$ in the setting of 
Fig.~\ref{fig.average}. \text{Bottom:} In order to study the occupation of the 
lattice site $x=-3$, one can introduce an occupation  observable. It alternates 
between two states, namely the particle is ``on $x$'' or ``not on $x$.'' The 
site occupation time $\Ot{x}{t}$ is then an accumulation of those waiting times 
that were spent ``on $x$.'' The stretches of time spent in the ``not on 
$x$''-state are essentially first-return times of the CTRW.}
 \label{fig.occupation}
\end{figure}

\section{Occupation time statistics}\label{sec.otf}
Let $f(\ot{x}{t};t)$ denote the PDF of the occupation time $\Ot{x}{t}$ and 
define its double Laplace transform $\trafo{f}(u;s)=\int_0^\infty\int_0^\infty 
e^{-st-u\ot{x}{t}}f(\ot{x}{t};t)\,dt\,d\ot{x}{t}$. To calculate it, we employ a 
method developed for the non-ergodic CTRW regime, i.e. 
$0<\alpha<1$~\cite{Bel2005,Bel2006}. It exploits the two-state renewal nature of 
the occupation process: the site $x$ alternates between being occupied or empty, 
see Fig.~\ref{fig.occupation}. To these two states we can associate sojourn time 
distributions. The time of continuous occupation of $x$ is distributed as 
$\psi(\tau)$. The time of continuous absence  from $x$ is essentially a 
first-return time: it measures how long it takes a CTRW particle to return to a 
site $x$ once it jumped off. We denote its PDF with $\psi_\subnx(\tau)$ 
(subscript: ``not on $x$''). Associated Laplace transforms are 
$\trafo{\psi}(s)=\int_0^\infty e^{-s\tau}\psi(\tau)\,d\tau$ and 
$\trafo{\psi}_\subnx(s)$ respectively. From these two distributions one can 
infer the statistics of the occupation time on $x$, see Eq.~(A13) of 
Ref.~\cite{Bel2006,footnote:initcond}:
\begin{align} 
  \trafo{f}(u;s)&=
  \frac{s\,\trafo{\psi}_\subnx(s)\left[1-\trafo{\psi}(s+u)\right]+(s+u)
  \left[1-\trafo{\psi}_\subnx(s)\right]}{s(s+u)\left[1-\trafo{\psi}
  (s+u)\trafo{\psi}_\subnx(s)\right]} .
  \label{ot}
\end{align}

We are interested in the large time statistics of occupation times, 
$t,\ot{x}{t}\gg\mean{\tau}$. In Laplace space, this corresponds to small $u,s$. 
For broad-tailed waiting time PDFs as in Eq.~\eqref{broadtail} with 
$1<\alpha<2$, the Laplace space asymptotics read
\begin{align}
 \trafo{\psi}(s) &\sim 1-\mean{\tau} s+As^\alpha +\Order{s^2} 
 \label{psiasymptotic} .
\end{align}
As demonstrated in Ref.~\cite{Bel2006}, one can use subordination arguments in 
the first-return time analysis to derive the respective small-$s$ expansion for 
the time of absence from $x$:
\begin{align}
 \trafo{\psi}_\subnx(s) &\sim 1-\mean{\tau^\suprw_\subnx}\big(\mean{\tau} s -A 
s^\alpha\big) +\Order{s^2} ,
 \label{psinxasymptotic}
\end{align}
where $\mean{\tau^\suprw_\subnx}$ is the mean first-return time of an ordinary 
random walk (RW). Here, ``ordinary'' means that the walker is subject to the 
same hopping probabilities $q_x$, but the waiting times on each lattice site are 
fixed to unity. The random walks studied in this work are recurrent (i.e. the 
ultimate return to any $x$ is certain), since they take place under the 
influence of a binding field $\pot{x}$. They also obey the detailed balance 
condition, Eq.~\eqref{detbal}. Under these circumstances, we have the simple 
relation~\cite{Bel2006,Benichou2014,footnote:mfpt}
\begin{equation}
 \mean{\tau_\subnx^\suprw}= \frac{1-\peq{x}}{\peq{x}} .
 \label{meantaurwnx}
\end{equation}

In order to obtain an approximation for the occupation time PDF in the limit of 
large times, we now insert the expansions in Eqs.~\eqref{psiasymptotic} 
and~\eqref{psinxasymptotic} into the exact  Eq.~\eqref{ot}; we drop, separately 
in the numerator and the denominator, anything beyond the two lowest order terms 
in $s$ and $(s+u)$, respectively. This gives
\begin{equation}
\hspace*{-5pt}
 \trafo{f}(u;s)\sim 
 \frac{1-C\left[\peq{x}(s+u)^{\alpha-1}+(1-\peq{x})s^{\alpha-1}\right]}%
 {s+u\peq{x}-C\left[\peq{x}(s+u)^{\alpha}+(1-\peq{x})s^{\alpha}\right]} 
 \label{ot:pdf:complete:trafo}
\end{equation}
for small, comparable values of $u$, $s$. The ratio $C=A/\mean{\tau}$ quantifies 
the relative width of the waiting time PDF. By fixing $s$ while expanding at 
$u\approx0$, we further get $\trafo{f}(u;s)\sim 
s^{-1}+u\peq{x}s^{-2}+\Order{u^2}$. We thus check that the 
PDF~\eqref{ot:pdf:complete:trafo} is properly normalized, its mean is indeed the 
ergodic limit, $\mean{\Ot{x}{t}}\sim t\peq{x}$, 
and its variance is finite, $\mean{[\Ot{x}{t}]^2}<\infty$.  The latter is 
natural, since occupation times are bounded, $0\leq\Ot{x}{t}\leq t$. 

We can now turn our attention to the fluctuations of the occupation times, 
$\Otf{x}{t}$. According to their definition~\eqref{otf}, their PDF 
$g(\otf{x}{t};t)$, is related through $\trafo{g}(k;s) = 
\int_{-\infty}^\infty\int_0^\infty 
e^{-st+ik\otf{x}{t}}g(\otf{x}{t};t)\,dt\,d(\otf{x}{t}) = 
\trafo{f}(-ik;s+ik\peq{x})$. Hence,
\begin{align}
 &\trafo{g}(k;s) \sim \label{otf:pdf:complete:trafo}\nn
 &\frac{1-C\left\{\peq{x}[s-ik(1-\peq{x})]^{\alpha-1}+(1-\peq{x})(s+ik\peq{x})^{
 \alpha-1}\right\}}
 {s-C\left\{\peq{x}[s-ik(1-\peq{x})]^{\alpha}+(1-\peq{x})(s+ik\peq{x})^{\alpha}
 \right\}} .
\end{align}
We would like to elaborate some more on the nature of this approximation. We 
have derived it under the premise that we are dealing with long measurement 
times $t$, hence we took $s$ to be small. Occupation times $\Ot{x}{t}$ typically 
grow with time, so we assumed $|k|$ is small and ``comparable'' to $s$. To be 
precise, the asymptotic approximation in Eq.~\eqref{otf:pdf:complete:trafo} is 
meaningful when $s$ is small and of the same order as $|k|^\alpha$, or $|k|$, or 
anything between. This corresponds exactly to the scaling limits in 
Eqs.~\eqref{otf:pdf:bulk:scalinglimit} and~\eqref{otf:pdf:ldev:scalinglimit}, 
where the fluctuations $\Otf{x}{t}$ are assumed to scale with $t^{1/\alpha}$ or 
$t$, respectively. We thus find our heuristic arguments in 
Sec.~\ref{sec.duality} being compatible with Eq.~\eqref{otf:pdf:complete:trafo}. 
This equation encodes the complete information on the asymptotic distribution of 
occupation times, but it lives in the abstract Fourier-Laplace space. Additional 
efforts are required to extract any applicable results. Therefore, we split the 
problem into two aspects, each focusing on one specific time scaling. This 
greatly simplifies and clarifies the picture.

\section{The stable law of bulk fluctuations}\label{sec.otf:bulk}
In order to find the precise bulk scaling function $g_\subbulk(\cdot)$, we 
continue from Eq.~\eqref{otf:pdf:complete:trafo}. The Fourier-Laplace space 
analogue to the scaling limit~\eqref{otf:pdf:bulk:scalinglimit} is to let $s$ 
become small, while fixing the ratio $s/|k|^\alpha$. 
Equation~\eqref{otf:pdf:complete:trafo} by this simplifies and becomes
\begin{align}
 &\trafo{g}(k;s) \sim \trafo{g}_\subbulk(k;s)= \nn
 &\frac{1}{s}\left\{ 1-C\left[(1-\peq{x})^{\alpha}\peq{x}\frac{(-ik)^\alpha}{s} 
+ (1-\peq{x})\peq{x}^{\alpha}\frac{(ik)^\alpha}{s}\right] \right\}^{-1}
\end{align}
The neglected higher order terms are of the order $\Order{s^{-1/\alpha}}$. To 
continue, we write
\begin{equation}
 (\pm ik)^\alpha=|k|^\alpha\left[\cos(\pi\alpha/2) 
 \pm i\sgn(k)\sin(\pi\alpha/2)\right] ,
\end{equation}
choosing principal values for the exponentiation and using 
$k,\alpha\in\mathbb{R}$. With this, the bulk statistics can be cast into the 
more compact form
\begin{align}
 \label{otf:pdf:bulk:trafo}
 \trafo{g}_\subbulk(k;s) &=
 \frac{1}{s-\kappa^\alpha|k|^\alpha[1-i\beta\,\sgn(k)\tan(\alpha\pi/2)]} ,
\end{align}
with parameters
\begin{subequations}
\label{otf:pdf:bulk:asymptotic}
\begin{align}
 \kappa &= \left\{C|\cos(\pi\alpha/2)| \left[(1-\peq{x})^\alpha\peq{x}
  +(1-\peq{x})\peq{x}^\alpha\right]\right\}^{1/\alpha} \\
 \beta&= 
 \frac{(1-\peq{x})^\alpha\peq{x}-(1-\peq{x})\peq{x}^\alpha}
  {(1-\peq{x})^\alpha\peq{x}+(1-\peq{x})\peq{x}^\alpha} .
\end{align}
Fourier-Laplace inversion of Eq.~\eqref{otf:pdf:bulk:trafo} yields the 
sought-after PDF $g_\subbulk(\otf{x}{t};t)$. It is now evident that we can write 
it in the scaling form
\begin{align}
  g_\subbulk(\otf{x}{t};t) 
  \equiv \frac{1}{t^{1/\alpha}} 
g_\subbulk\left(\frac{\otf{x}{t}}{t^{1/\alpha}}\right)
  = \frac{1}{\kappa 
t^{1/\alpha}}\ell_{\alpha,\beta}\left(\frac{\otf{x}{t}}{\kappa 
t^{1/\alpha}}\right) .
\end{align}
Thus, at fixed $t$, the scale parameter $\kappa$ measures the bulk PDF's width. 
For any fixed waiting time parameters, $\kappa$ has a maximum at $\peq{x}=1/2$ 
and vanishes if $\peq{x}$ tends to either $0$ or $1$. Moreover, the 
characteristic function
 \begin{align}
  \label{otf:pdf:bulk:asymptotic:scalingfct}
  \trafo{\ell}_{\alpha,\beta}(k) 
  &= \int_{-\infty}^\infty e^{iky} \ell_{\alpha,\beta}(y)\,dy \nn
  &= \exp\left\{-|k|^\alpha[1-i\beta\,\sgn(k)\tan(\alpha\pi/2)]\right\} ,  
 \end{align}
\end{subequations}
defines the $\alpha$-stable law~\cite{GCLT,StableProc} 
$\ell_{\alpha,\beta}(y)$. 

Following the intuitive reasoning along the lines of the generalized limit 
theorem (see Sec.~\ref{sec.duality}), the basic $\alpha$-stable form of the 
scaling function was to be expected. Nonetheless,  some of its specific features 
are remarkable. The skewness parameter $-1<\beta<1$ indicates an asymmetry; see 
the sample plots in Fig.~\ref{fig.otf.pdf}. This contrasts the familiar Gaussian 
fluctuations, which appear, e.g. in the ordinary random walk. Also, in other 
physical systems where L\'evy-stable statistics play a role, one encounters 
typically rather $\beta=0$ or $\beta=\pm1$. The asymmetry is controlled through 
$\alpha$ and $\peq{x}$. A symmetric distribution is obtained when $\peq{x}=1/2$, 
or in the limit $\alpha=2$, which generally recovers Gaussianity.

Another surprising feature of Eq.~\eqref{otf:pdf:bulk:asymptotic} is its 
independence of any $\peq{x'}$ with $x'\neq x$. Naturally, at large times, the 
average occupation time becomes $\mean{\Ot{x}{t}}\sim t\peq{x}$ by virtue of 
ergodicity. But even more, the complete PDF of fluctuations $\Otf{x}{t}$ is 
controlled by the single steady-state probability $P_x$, in conjunction with the 
waiting time characteristics $\alpha$ and $C$.

One thing appears to be odd though. Because of $0\leq\Ot{x}{t}\leq t$, the  
fluctuations have natural bounds,
\begin{equation}
 -t\peq{x}\leq\Otf{x}{t}\leq t(1-\peq{x}) .
 \label{otf:bounds}
\end{equation}
But the support of the stable law $g_\subbulk(\cdot)$ is the complete real line. 
More seriously, $g_\subbulk(\cdot)$ takes over the broad 
tails~\eqref{broadtail}, i.e. $g_\subbulk(y)\simeq y^{-\alpha-1}$ for 
large~$y$~\cite{GCLT,StableProc}. Hence, high-order moments diverge. This 
includes even the variance, 
$\mean{[\Otf{x}{t}]^2}_\subbulk=t^{2/\alpha}\int_{-\infty}^\infty 
y^2g_\subbulk(y)\,dy=\infty$, easily the most common quantitative measure for 
fluctuations in experimental or numeric data. Evidently, in the case of CTRW, by 
focusing on the bulk scaling~\eqref{otf:pdf:bulk:scalinglimit}, we are missing 
vital statistical information. To complete the picture, we need to explicitly 
consider the rare large fluctuations.

\section{The infinite density of large fluctuations}\label{sec.otf:ldev}
The nonstandard scaling limit~\eqref{otf:pdf:ldev:scalinglimit} can be obtained 
as follows. According to Eq.~\eqref{otf:bounds}, the largest occupation time 
fluctuations $\Otf{x}{t}$ are of the order of the measurement time $t$ itself. 
We therefore restart from Eq.~\eqref{otf:pdf:complete:trafo}, let again $s$ be 
small, but now fix the ratio $s/|k|$.
This gives
\begin{align}
  &\trafo{g}(k;s) \sim \trafo{g}_\subldev(k;s) = \frac{1}{s}+ 
\frac{C}{s^{2-\alpha}}\times \nn
  &\begin{aligned}
    & \;\; \Bigg\{
      \peq{x}\left[1-(1-\peq{x})\frac{ik}{s}\right]^{\alpha} 
+(1-\peq{x})\left[1+\peq{x}\frac{ik}{s}\right]^{\alpha} \\
    & \;\; 
      -\peq{x}\left[1-(1-\peq{x})\frac{ik}{s}\right]^{\alpha-1} 
-(1-\peq{x})\left[1+\peq{x}\frac{ik}{s}\right]^{\alpha-1} \Bigg\} \\
    &= \frac{1}{s} +  
C\peq{x}\;\trafo{R}_\alpha\big((1-\peq{x})k;s\big)+C(1-\peq{x})\; 
\trafo{R}_\alpha\big({-\peq{x}}k;s\big) .
  \end{aligned}
  \label{otf:pdf:ldev:trafo}
\end{align}
In the first step, we dropped all terms which are of  
$\Order{s^{-1+2(\alpha-1)}}$. In the second, we made some simple rearrangements 
and introduced the auxiliary function
\begin{align}
 \trafo{R}_\alpha(qk;s) &=\frac{-ikq}{s^2}(s-ikq)^{\alpha-1} \nn
 &=  \left[ -\frac{ikq}{s}+\frac{(ikq)^2}{s^2}\right] (s-ikq)^{\alpha-2}
 .
 \label{otf:pdf:ldev:rdef}
\end{align}
We show below that the Fourier-Laplace inversion of $\trafo{R}_\alpha$, and 
hence of Eq.~\eqref{otf:pdf:ldev:trafo} yields the scaling function 
$g_\subldev(\cdot)$ appearing in Eq.~\eqref{otf:pdf:ldev:scalinglimit}.

But before we start, some technical remarks are in place. In 
Eq.~\eqref{otf:pdf:ldev:trafo}, the term $1/s$ is, strictly speaking, the 
leading order term in this small-$s$ expansion. It becomes $\delta(\otf{x}{t})$ 
upon Fourier-Laplace inversion, i.e. a Dirac $\delta$-distribution. This does 
not come as a surprise. We are now rescaling the statistics linearly with time, 
which is faster than the bulk scaling $\Otf{x}{t}\sim t^{1/\alpha}$, $1<\alpha$. 
Hence, we ``squeeze'' the bulk realizations onto the origin. The higher order 
terms in Eq.~\eqref{otf:pdf:ldev:trafo} contain exactly the information on those 
fluctuations that deviate from the bulk behavior. Hence, we suppress the 
$\delta(\otf{x}{t})$-term in the oncoming calculations.  However, notice that 
$\delta(\otf{x}{t})$ is by itself a PDF. Removing it from the equation naturally 
causes some issues with normalization for the remaining parts. Therefore, we 
should not interpret the $g_\subldev(\cdot)$ literally as a PDF. Mathematically, 
we should rather treat it as a generalized function. When calculating an average 
with it as $\mean{\phi(\Otf{x}{t})}_\subldev=\int g_\subldev(\otf{x}{t};t) 
\phi(\otf{x}{t}) d(\otf{x}{t})$, then we make it act on a test function $\phi$. 
Not all test functions need to be ``suitable'', i.e. integrable, with respect to 
the infinite density.
In Sec.~\ref{sec.moments}, we elaborate more on this issue.

We now invert the $\trafo{R}_\alpha(qk;s)$ as defined in 
Eq.~\eqref{otf:pdf:ldev:rdef}. We proceed in two steps: first, we go from 
Laplace space to time, $s\rightarrow t$; second, from Fourier space to 
fluctuations, $k\rightarrow\otf{x}{t}$. We write the associated transformed 
functions as $\trafo{R}_\alpha(qk;s)=\int_0^\infty e^{-st} 
\trafoF{R}_\alpha(qk;t)\,dt$ and $\trafoF{R}_\alpha(qk;t)=\int_{-\infty}^\infty 
e^{ik\otf{x}{t}} R_\alpha(\otf{x}{t}/q;t)\,d(\otf{x}{t})/q$. If we start from 
the second line of Eq.~\eqref{otf:pdf:ldev:rdef}, then the first step, the 
Laplace inversion, $s\rightarrow t$, becomes
\begin{align}
 \trafoF{R}_\alpha(qk;t) &= -ikq \int_0^t e^{ikq\omega} 
\frac{\omega^{-(\alpha-1)}}{\Gamma(2-\alpha)}\,d\omega \nn
 &\quad +(ikq)^2 \int_0^t  \,e^{ikq\omega} 
(t-\omega)\frac{\omega^{-(\alpha-1)}}{\Gamma(2-\alpha)}\,d\omega .
 \label{ldev:rF}
\end{align}
We can put this in a more elegant form, which makes explicit the time scaling 
nature of the function:
\begin{align}
 \trafoF{R}_\alpha(qk;t) = \frac{\alpha}{t^{\alpha-1}|\Gamma(1-\alpha)|} \, 
\trafoF{\infd}(\zeta) , \qquad  \zeta &= kqt ,
 \label{otf:pdf:ldev:rscalingfunction}
\end{align}
where
\begin{align}
 \trafoF{\infd}(\zeta) &=\frac{1}{\alpha(\alpha-1)} \Big[-i\zeta \int_0^1 
e^{i\zeta\omega} \omega^{-(\alpha-1)}\,d\omega \nn
 &\quad\quad +(i\zeta)^2 \int_0^1  \,e^{i\zeta\omega}  
(1-\omega)\,\omega^{-(\alpha-1)}\,d\omega \Big] .
 \label{otf:pdf:ldev:scalingfunctionF}
\end{align}
The remaining task is the Fourier inversion $k\rightarrow\otf{x}{t}$, or 
respectively, $\zeta\rightarrow z=\otf{x}{t}/(qt)$. For this, we translate 
powers of $\zeta$ into derivatives $\partial/\partial z$ and the exponential 
$e^{i\zeta\omega}$ into $\delta(z-\omega)$. We thus get
the inverse Fourier-Transform of $\trafoF{\infd}(\zeta)$,
\begin{align}
 \infdens{z} 
   &=\frac{1}{\alpha(\alpha-1)} \left\{ \frac{\partial}{\partial z}\int_0^1 
\delta(z-\omega) \,\omega^{-(\alpha-1)}\,d\omega \right. \nn
   &\hspace{12mm} \left.+\frac{\partial^2}{\partial z^2} \int_0^1 
\delta(z-\omega)\left[(1-\omega)\,\omega^{-(\alpha-1)}\right]\,d\omega\right\} 
\nn
   &=\frac{1}{\alpha(\alpha-1)} \left\{ \frac{\partial}{\partial z} 
\left[\indfu{0<z\leq 1} z^{-(\alpha-1)}\right] \right.\nn
   &\hspace{12mm}\left. +\frac{\partial^2}{\partial z^2}\left[ \indfu{0<z\leq 1} 
(1-z)\,z^{-(\alpha-1)}  \right]\right\}
 .\label{otf:pdf:ldev:scalingfunction:differentiate}
\end{align}
The indicator function $\indfu{\cdot}$ is $1$ when the condition in the argument 
is fulfilled, and $0$ otherwise. Hence, it is a discontinuous step function, and 
its differentiation gives rise to several $\delta$-distributions in the 
expression. For reasons that we elaborated on above, we drop the contributions 
from the boundary $z=0$. In App.~\ref{appsec.otf:pdf:ldev}, we show that the 
other peaks at $z=1$ have zero contribution. Performing the differentiations in 
Eq.~\eqref{otf:pdf:ldev:scalingfunction:differentiate} therefore gives
\begin{subequations}
\begin{align}
 \infdens{z}
 &=\indfu{0<z\leq1} z^{-1-\alpha} \left(1-\frac{\alpha-1}{\alpha}z\right) .
 \label{otf:pdf:ldev:asymptotic:scalingfct}
\end{align}
Tracing back the definitions in Eqs.~\eqref{otf:pdf:ldev:scalingfunctionF}, 
\eqref{otf:pdf:ldev:rscalingfunction} and~\eqref{otf:pdf:ldev:trafo}, we finally 
arrive at
 \label{otf:pdf:ldev:asymptotic}
 \begin{align}
  &g_\subldev(\otf{x}{t};t) \equiv \frac{1}{t^\alpha} 
g_\subldev\!\left(\frac{\otf{x}{t}}{t}\right) =
  \frac{C\alpha }{t^{\alpha}|\Gamma(1-\alpha)|} \times \\
  &\quad\begin{dcases}
    \frac{\peq{x}}{1-\peq{x}}\,\infdens{\frac{\otf{x}{t}}{t(1-\peq{x})}}
    ,& 0<\frac{\otf{x}{t}}{t}\leq1-\peq{x} ,\\
    \frac{1-\peq{x}}{\peq{x}}\,\infdens{\frac{|\otf{x}{t}|}{t\peq{x}}}
    ,& -\peq{x}\leq\frac{\otf{x}{t}}{t}<0 ,\\
    0 ,&\text{otherwise} .
   \end{dcases}\nonumber
 \end{align}
\end{subequations}
This scaling function is defined within the appropriate 
bounds~\eqref{otf:bounds} and correctly predicts the behavior of the large 
occupation times, see Fig.~\ref{fig.otf.pdf}. Again, the expression depends 
solely on $\alpha$, $C$, and the single equilibrium probability $\peq{x}$, and 
it is symmetric only if $\peq{x}=1/2$. 

Due to the fast divergence $\simeq|\otf{x}{t}|^{-\alpha-1}$ at the origin, 
$g_\subldev(\cdot)$ is not normalizable. This is why we call it an infinite 
density. This distinguished property should not be a cause of concern. Of 
course, at any finite time $t$, the exact PDF $g(\cdot)$ is perfectly 
normalized. Its central part is best approximated by the stable law 
$g_\subbulk(\cdot)$, Eq.~\eqref{otf:pdf:bulk:asymptotic}, while large 
fluctuations are best studied in terms of $g_\subldev(\cdot)$, 
Eq.~\eqref{otf:pdf:ldev:asymptotic}. In particular, the pole of the infinite 
density is completely congruent with the tails of the $\alpha$-stable bulk 
distribution~\cite{StableProc}: Eqs.~\eqref{otf:pdf:bulk:scalinglimit}, 
\eqref{otf:pdf:ldev:scalinglimit}, \eqref{otf:pdf:bulk:asymptotic}, and 
\eqref{otf:pdf:ldev:asymptotic} consistently yield 
\begin{subequations}
\begin{align}
 g(\otf{x}{t};t) 
 &\sim 
\frac{1}{t^{1/\alpha}}g_\subbulk\left(\frac{\otf{x}{t}}{t^{1/\alpha}}\right) 
 \sim \frac{1}{t^\alpha}g_\subldev\left(\frac{\otf{x}{t}}{t}\right) \nn
 &\sim \frac{C\alpha (1-\peq{x})^\alpha\peq{x} t}{|\Gamma(1-\alpha)|} 
{\otf{x}{t}}^{-\alpha-1} 
 \nn &\quad\text{ for }\kappa t^{1/\alpha}\ll\otf{x}{t}\ll t(1-\peq{x}),\\
 g(\otf{x}{t};t) 
 &\sim 
\frac{1}{t^{1/\alpha}}g_\subbulk\left(\frac{\otf{x}{t}}{t^{1/\alpha}}\right) 
 \sim \frac{1}{t^\alpha}g_\subldev\left(\frac{\otf{x}{t}}{t}\right) \nn
 &\sim \frac{C\alpha (1-\peq{x})\peq{x}^\alpha t}{|\Gamma(1-\alpha)|} 
|\otf{x}{t}|^{-\alpha-1} 
 \nn &\quad\text{ for } -t\peq{x}\ll\otf{x}{t}\ll -\kappa t^{1/\alpha}. 
\end{align}
\label{otf:pdf:powerlaw}%
\end{subequations}
The power law $|\otf{x}{t}|^{-\alpha-1}$ is responsible for both types of 
unphysical divergences in the asymptotics: the divergent moments of the stable 
laws~\eqref{otf:pdf:bulk:asymptotic} and the divergent norm of the infinite 
density~\eqref{otf:pdf:ldev:asymptotic}. But by patching the two asymptotic 
expressions along the common power law overlap we obtain a complete, consistent 
statistical long-time description, bare of any such dubious divergences.

\section{Ensemble averages}\label{sec.moments}
When calculating an average of the form $\mean{\phi(\Otf{x}{t})}$, we are 
confronted with the question of how to apply the above asymptotic results. Is 
the average obtained from the L\'evy-stable bulk statistics $g_\subbulk(\cdot)$, 
Eq.~\eqref{otf:pdf:bulk:asymptotic}? Or is it sensitive to the large 
fluctuations, so that we need the infinite density $g_\subldev(\cdot)$, 
Eq.~\eqref{otf:pdf:ldev:asymptotic}? Or may we even have to consider the 
complete, yet implicit solution $\trafo{g}(\cdot)$, 
Eq.~\eqref{otf:pdf:complete:trafo}? In general, we find that $g_\subbulk(\cdot)$ 
and $g_\subldev(\cdot)$ complement one another and yield the entire information 
on the long-time asymptotics. This becomes most evident in the study of absolute 
$q$th order moments, $\phi(\Otf{x}{t})=|\Otf{x}{t}|^q$. We show in 
App.~\ref{appsec.otf:moments} that, with the exception of $q=\alpha$, the 
moments fall into two distinct classes, which are exclusively integrable with 
respect to either the stable law or the infinite density: 
\begin{subequations}
\label{otf:moments}
\begin{equation}
 \mean{|\Otf{x}{t}\!|^q}  \sim
  \begin{dcases}
   \mean{|\Otf{x}{t}\!|^q}_\subbulk=\mql \,t^{q/\alpha} ,& q<\alpha ,\\
   \mean{|\Otf{x}{t}\!|^q}_\subldev=\mqg \,t^{q+1-\alpha} ,& q>\alpha ,
  \end{dcases}
  \label{otf:moments:scaling}
 \end{equation}
where
 \begin{align}
  \mql &= \kappa^q \int_{-\infty}^\infty |y|^q \ell_{\alpha,\beta}(y) \,dy    
  \quad\text{ [see Ref.~\cite{StableProc}, chap.~1.2]}
  ,\nn
  \mqg 
  &= \frac{C\alpha\left[ \peq{x}(1-\peq{x})^q   +(1-\peq{x}){\peq{x}}^q  
\right]}{|\Gamma(1-\alpha)|}  \int^1_0 z^q\,\infdens{z} \,dz \nn
  &= \frac{Cq\left[ \peq{x}(1-\peq{x})^q   +(1-\peq{x}){\peq{x}}^q  
\right]}{|\Gamma(1-\alpha)|(q-\alpha)(q-\alpha+1)} .
  \label{otf:moments:coeff}
 \end{align}
\end{subequations}
We stress that the coefficients $\mql$ and $\mqg$ are finite in their respective 
parameter domains. The stable laws $\ell_{\alpha,\beta}(y)$ do possess finite 
moments of order $q<\alpha$. The expression $z^q\,\infdens{z}$ is perfectly 
integrable when $q>\alpha$. We can thus use, in this context, the infinite 
density $g_\subldev(\cdot)$ as if it was an ordinary PDF -- despite the pole at 
the origin. We add that the borderline case $\mean{|\Otf{x}{t}|^\alpha}\simeq 
t\ln t$ is obtained by carefully accounting for contributions from both the bulk 
and large fluctuations, see App.~\ref{appsec.otf:moments}.

According to Eq.~\eqref{otf:moments:scaling}, the different time scalings of 
bulk versus large fluctuations manifest in a dual scaling of moments. The same 
duality has been discussed with respect to strong anomalous 
diffusion~\cite{Vulpiani1999,Rebenshtok2014}.
In particular, we have 
$\mean{[\Otf{x}{t}]^2}\sim\mean{[\Otf{x}{t}]^2}_\subldev\simeq t^{3-\alpha}$. 
Even the variance -- usually not perceived as a high-order moment -- does probe 
the statistics of the large fluctuations. Its increase with time is faster than 
anticipated from the bulk scaling ($[\Otf{x}{t}]^2\sim t^{2/\alpha}$).

\section{Time-averaged observables}\label{sec.taobs}
In this section, we show that the connection between dual scaling laws and 
stable and infinite densities is not specific to occupation time statistics, but 
extends to a larger class of observables. Assume an observable $\Obs$ takes on 
the value $\obs_x$ when the random walker sits on site $x$. Then, by virtue of 
ergodicity, Eq.~\eqref{ergodic}, the time average
\begin{align}
 \ta{\Obs}(t) &= \frac{1}{t} \sum_x \obs_x \Ot{x}{t}
 \label{taobs}
 \intertext{converges to the equilibrium ensemble average}
 \meaneq{\Obs}&= \sum_x \obs_x \peq{x}
\end{align}
in the limit $t\rightarrow\infty$. At finite time $t$, we define the 
fluctuations of the time average as
\begin{equation}
 \ta{\Obsf}(t)=\ta{\Obs}(t)-\meaneq{\Obs} =\frac{1}{t}\sum_x \obs_x \Otf{x}{t}
 .
 \label{taobsf}
\end{equation}
We denote the respective PDF by $h(\ta{\obsf};t)$. For reasons which become 
apparent below, we restrict ourselves to observables and systems where the 
equilibrium average $\meaneq{|\Obsf|^\alpha} = \sum_x 
|\obs_x-\meaneq{\Obs}|^\alpha \peq{x}$ is finite.

The general time average is defined through Eq.~\eqref{taobs} in terms of the 
occupation time fractions $\Ot{x}{t}/t$. Conversely, the occupation time 
fraction of a specific region $\domain$ of the lattice is proportional to the 
time average of a particular observable. Namely, we can define an occupation 
observable $\Obs$ through an indicator function $\indfu{\cdot}$ :
\begin{subequations}
\label{taobs:otf}
\begin{equation}
 \obs_x= \indfu{x\in \domain}=
 \begin{cases}
  1 & \text{for }x\in \domain,\\
  0 & \text{for }x\notin \domain.
 \end{cases}
 \label{taobs:otf:def}
\end{equation}
The time average of an occupation observable can be written as
\begin{equation}
 \ta{\Obs}(t)= \frac{1}{t}\sum_{x} \indfu{x\in \domain} \Ot{x}{t} = 
\frac{1}{t}\sum_{x\in \domain} \Ot{x}{t} = \frac{\Ot{\domain}{t}}{t} ,
\end{equation}
and likewise
\begin{equation}
 \ta{\Obsf}(t)= \frac{\Otf{\domain}{t}}{t} .
\end{equation}
\end{subequations}
In the simplest case, $\domain$ would only consist of a single lattice point $x$ 
(see, e.g., Fig.~\ref{fig.occupation}, where $x=-3$). We then return to the 
scenario of the previous sections. In this sense, the single-site occupation 
time fractions $\Ot{x}{t}/t$ are special cases of the general time average 
$\ta{\Obs}(t)$ in Eq.~\eqref{taobs}. We are hence now extending the discussion 
to a broader class of observables. The detailed calculus is similar to the 
previous and can be found in App.~\ref{appsec.taobs}. We focus on the implied 
generalizations for the dual time scaling, the bulk and the large fluctuation 
laws.

\subsection{The dual time scaling of time averages}\label{sec.taobs:duality}
The fluctuations of a general time average $\ta{\Obsf}(t)$ should, according to 
definition~\eqref{taobsf}, have essentially the same time scaling behavior as 
the fluctuations of the occupation time fractions $\Otf{x}{t}/t$. That is, for 
bulk fluctuations we expect that $\ta{\Obsf}(t)\sim t^{1/\alpha}/t 
=t^{-(\alpha-1)/\alpha}$, while the large fluctuations should be of constant 
order. Indeed, we find the bulk behavior
\begin{align}
  \label{taobs:pdf:bulk:scalinglimit}
  h(\ta{\obsf};t) &\sim t^{(\alpha-1)/\alpha}\,
   h_\subbulk\left(\ta{\obsf}\,t^{(\alpha-1)/\alpha}\right) ,\\
\intertext{and the large fluctuations scaling limit is}
  \label{taobs:pdf:ldev:scalinglimit}
  h(\ta{\obsf};t) &\sim 
  \frac{1}{t^{\alpha-1}}h_\subldev\left(\ta{\obsf}\right) .
\end{align}
Equations~\eqref{otf:pdf:bulk:scalinglimit} 
and~\eqref{otf:pdf:ldev:scalinglimit} are to be viewed as special cases. This 
basic nature of the dual time scaling is thus universal for all time averages, 
including occupation times. What remains to be investigated is the exact form of 
the scaling functions $h_\subbulk(\cdot)$ and $h_\subldev(\cdot)$ and their 
sensitivity to the choice of the observable $\Obs$.

\subsection{The stable law of bulk fluctuations}\label{sec.taobs:bulk}
The bulk scaling limit for large $t$ reads
\begin{subequations}
 \label{taobs:pdf:bulk:asymptotic}
 \begin{align}
  h_\subbulk(\ta{\obsf};t)
  &\equiv t^{(\alpha-1)/\alpha}\, 
   h_\subbulk\left(\ta{\obsf}\,t^{(\alpha-1)/\alpha}\right) \nn 
  &= \frac{t^{(\alpha-1)/\alpha}}{\kappa}\ell_{\alpha,\beta}
   \left(\frac{\ta{\obsf}\, t^{(\alpha-1)/\alpha}}{\kappa}\right) ,
 \end{align}
 where now
 \begin{align}
  \kappa &= \left[C|\cos(\pi\alpha/2)|\meaneq{|\Obsf|^\alpha}\right]^{1/\alpha},
 \end{align}
 and the $\alpha$-stable law $\ell_{\alpha,\beta}(y)$ is as in 
Eq.~\eqref{otf:pdf:bulk:asymptotic:scalingfct}, only with $\beta$ defined as
 \begin{align}
  \beta&= \frac{\meaneqcond{|\Obsf|^\alpha}{\Obsf>0}
   -\meaneqcond{|\Obsf|^\alpha}{\Obsf<0}}{\meaneq{|\Obsf|^\alpha}} .
 \end{align}
\end{subequations}
We have introduced here the $\alpha$th order absolute moments of equilibrium 
fluctuations
\begin{subequations}
\label{taobs:eqmom}
\begin{align}
 \meaneq{|\Obsf|^\alpha} 
 &= \meaneq{\left|\Obs-\meaneq{\Obs}\right|^\alpha} \nn
 &= \sum_x \left|\obs_x-\meaneq{\Obs}\right|^\alpha \peq{x}
\intertext{and respective constrained averages}
 \meaneqcond{|\Obsf|^\alpha}{\Obsf\gtrless z} 
 &= \meaneq{\indfu{\Obs -\meaneq{\Obs} \gtrless z}\left|\Obs 
-\meaneq{\Obs}\right|^\alpha} \nn
 &= {\sum_x}^\gtrless |\obs_x-\meaneq{\Obs}|^\alpha \peq{x} .
\end{align}
\end{subequations}
The constrained sums $\sum^\gtrless_x$ run over all $x$ where $\obs_x\gtrless 
z+\meaneq{\Obs}$, respectively.

The main observation here is that the basic $\alpha$-stable character of bulk 
fluctuations is universal for all observables. Still, the details are 
observable-specific, in particular the skewness parameter $\beta$. If the 
observable $\Obs$ in equilibrium symmetrically fluctuates around its mean, then 
$\beta=0$, so the fluctuations of the time average are also symmetric. The 
converse, however, does not hold. One can think of a situation where $\Obs$ is 
asymmetric in equilibrium, but in such a way that we happen to have 
$\meaneqcond{|\Obsf|^\alpha}{\Obsf<0}=\meaneqcond{|\Obsf|^\alpha}{\Obsf>0}$, so 
that the time average is symmetric. Thus, there seems to be no simple one-to-one 
correspondence between the symmetry of an observable in equilibrium and the 
symmetry of the respective time average -- that is, with respect to the bulk 
statistics.

\subsection{The infinite density of large fluctuations}\label{sec.taobs:ldev}
The scaling function in the large fluctuations 
limit~\eqref{taobs:pdf:ldev:scalinglimit} is given through
\begin{subequations}
 \label{taobs:pdf:ldev:asymptotic}
 \begin{align}
  \hspace{-3mm} 
  h_\subldev(\ta{\obsf};t) 
  &\equiv \frac{1}{t^{\alpha-1}}h_\subldev\left(\ta{\obsf}\right) \nn
  &=\frac{C\alpha }{t^{\alpha-1}|\Gamma(1-\alpha)|} 
  \begin{dcases}
    \infdensg{\ta{\obsf}}{\Obs}
    ,& 0<\ta{\obsf} ,\\
    \infdensl{\ta{\obsf}}{\Obs}
    ,& \ta{\obsf}<0 ,
   \end{dcases}
 \end{align}
introducing the infinite densities
 \begin{align}
  \infdensgl{z}{\Obs} &= |z|^{-\alpha-1} 
   \Big[\meaneqcond{|\Obsf|^\alpha}{\Obsf\gtrless z}  \nn
  &\qquad -  \frac{\alpha-1}{\alpha}
   |z|\meaneqcond{|\Obsf|^{\alpha-1}}{\Obsf\gtrless z} \Big] .
  \label{taobs:pdf:ldev:asymptotic:scalingfct}
 \end{align}
\end{subequations}
The non-integrable pole $\simeq \left|\ta{\obsf}\right|^{-\alpha-1}$ at the 
origin consistently overlaps with the broad tails of the $\alpha$-stable bulk 
PDF in Eq.~\eqref{taobs:pdf:bulk:asymptotic}. Therefore, this feature is again 
universal for all observables. But the specifics of the observable do determine 
the shape of the complete infinite density. We stress that the constrained 
moments that enter Eq.~\eqref{taobs:pdf:ldev:asymptotic} explicitly depend on 
$\ta{\obsf}$. Because of this, the analysis of the large fluctuations does 
reveal the symmetry and other characteristics of the observable $\Obs$. In 
contrast to the mere bulk statistics, the PDF $h_\subldev(\cdot)$ is symmetric 
if and only if the $\Obs$ fluctuates symmetrically around its mean in the 
equilibrium state.

\begin{figure*}
 \includegraphics[]{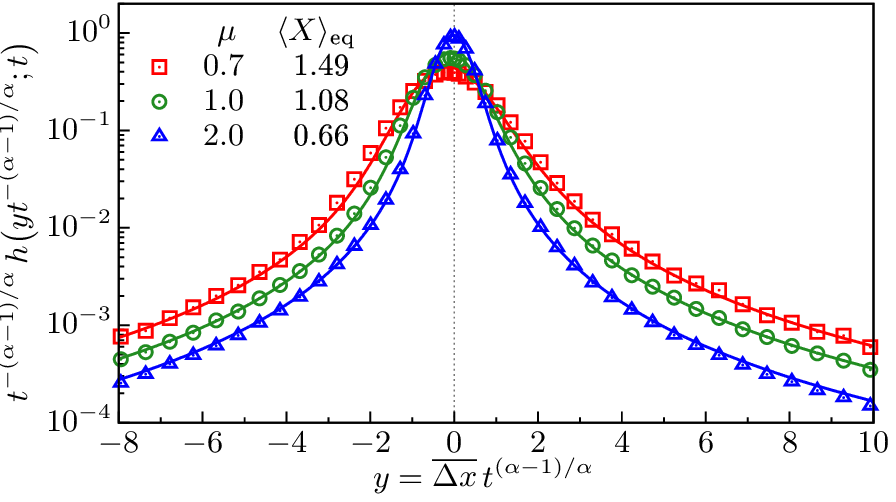}\includegraphics[]{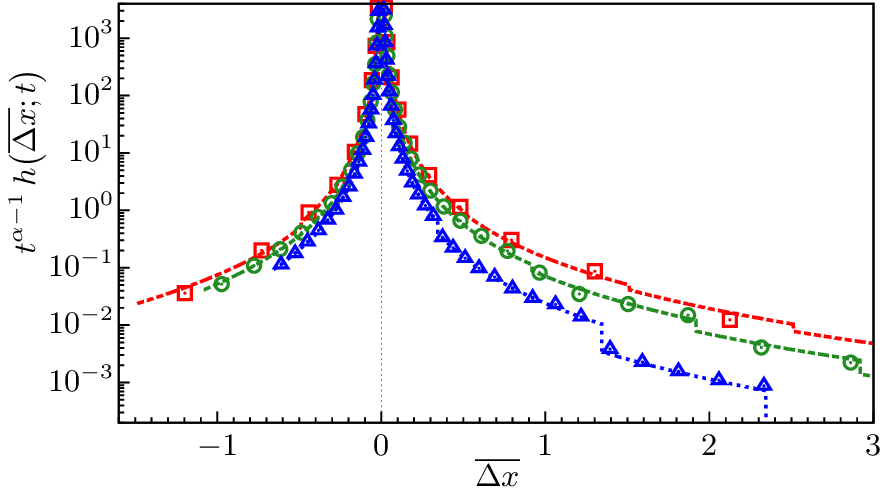}
 \caption{Scaled PDF of fluctuations $\ta{\Delta X}(t)$ of the time-averaged 
position for confined CTRW. The confinement is a constant bias to the left, 
$\pot{x}/(\kb\temp)=\mu x$, towards a reflecting boundary at the origin. 
Different symbols are simulation results for different bias strengths, $\mu=0.7$ 
($7\times10^5$ trajectories, $t=10^7$), $\mu=1.0$ ($6\times10^6$ trajectories, 
$t=10^6$) and $\mu=2.0$ ($2\times10^7$ trajectories, $t=10^5$). In all cases, 
waiting times were broad-tailed as in Eq.~\eqref{broadtail} with $\alpha=1.5$, 
$\mean{\tau}=1$ and $A=0.5$. Both panels depict the same data. \textit{Left:} 
Scaling with $t^{-(\alpha-1)/\alpha}$  as in 
Eq.~\eqref{taobs:pdf:bulk:scalinglimit} yields the asymmetric, stable bulk 
statistics $h_\subbulk(\cdot)$ of Eq.~\eqref{taobs:pdf:bulk:asymptotic} 
(continuous lines), with the equilibrium statistics given by 
Eqs.~\eqref{tax:mean} and~\eqref{tax:eqmom:constrained}.  \textit{Right:} The 
nonstandard time scaling in Eq.~\eqref{taobs:pdf:ldev:scalinglimit} produces the 
infinite density $h_\subldev(\cdot)$ for the large fluctuations as in 
Eq.~\eqref{taobs:pdf:ldev:asymptotic} (dashed lines). The infinite density is 
bounded only to the left due to the reflective wall, and its discontinuous steps 
reveal the discrete lattice structure. This information could not be extracted 
from the bulk plot on the left.
 }
 \label{fig.tax.pdf}
\end{figure*}

\subsection{Example: Time-averaged position of a particle biased towards a 
reflective wall}\label{sec.taobs:tax}
As an example, we study the time-averaged position,
\begin{align}
 \obs_x&=x, &\ta{\Obs}(t)&=\ta{X}(t),
\end{align}
for a thermal CTRW as defined in Sec.~\ref{sec.model}. We consider a constant 
force $-F<0$ as a bias, which drives the particle against a reflective wall at 
$x=0$. For convenience, we define a parameter $\mu=F/(\kb\temp)$  that 
quantifies the competition between the localizing bias $F$ and the diffusive 
spreading at finite temperature $\temp$. On our lattice we define the discrete 
potential landscape
\begin{align}
 \frac{\pot{x}}{\kb\temp}
 &=\begin{dcases}
   \infty ,& x\leq-1 ,\\
   \ln(1+e^\mu) ,& x=0 ,\\
   \mu x ,& x\geq1 ,
 \end{dcases}
\end{align}
which is linear in $x$ for all $x\geq1$. Thermal detailed balance, 
Eq.~\eqref{detbal}, then implies that 
\begin{align}
 q_x&=
 \begin{dcases}
  1 ,&x=0 ,\\
  \frac{1}{1+e^\mu}\equiv q ,& x\geq1. 
 \end{dcases}
\end{align}
Hence, all steps are taken preferably to the left with probability $1-q>1/2$; 
only at the origin, they are reflected to the right. The Boltzmann-Gibbs 
equilibrium PDF, Eq.~\eqref{boltzmann}, for these dynamics reads
\begin{align}
 \peq{x} 
 &=\frac{1}{Z}\begin{dcases}
   0 ,& x\leq-1 ,\\
   \frac{1}{1+e^\mu} ,& x=0 ,\\
   e^{-\mu x} ,& x\geq1 ,
 \end{dcases}
 \label{tax:peq}
\end{align}
with the partition function $Z=[\sinh(\mu)]^{-1}$.
The equilibrium position is given by
\begin{equation}
 \label{tax:mean}
 \meaneq{X}= 
 \sum_{x=0}^\infty x\peq{x}= \frac{e^\mu+1}{(e^\mu-1)^2} .
\end{equation}
For the statistics of the time average, we further need the absolute moments of 
Eq.~\eqref{taobs:eqmom}:
\begin{subequations}
\label{tax:eqmom:constrained}
\begin{align}
 \meaneqcond{|\Delta X|^\alpha}{\Delta X>z}  
 &= \sum_{x=\lceil\meaneq{X}+z\rceil}^\infty \left(x-\meaneq{X}\right)^\alpha 
\frac{e^{-\mu x}}{Z}
\end{align}
for all $z>0$. We used the notation $\lceil x\rceil$ to designate the smallest 
integer $\geq x$. Similarly, we can find, for all $-\meaneq{X}\leq z<0$,
\begin{align}
 \meaneqcond{|\Delta X|^\alpha}{\Delta X<z}
 &= \sum_{x=0}^{\lfloor\meaneq{X}+z\rfloor} \left(\meaneq{X}-x\right)^\alpha 
\frac{e^{-\mu x}}{Z} .
\end{align}
Here, $\lfloor x\rfloor$ is the largest integer $\leq x$. Of course we 
implicitly obtain
\begin{align}
 \meaneq{|\Delta X|^\alpha} &= \meaneqcond{|\Delta X|^\alpha}{\Delta X<0} + 
\meaneqcond{|\Delta X|^\alpha}{\Delta X>0} .
\end{align}
\end{subequations}

Equations~\eqref{tax:mean} and~\eqref{tax:eqmom:constrained} contain all the 
equilibrium statistics that we need to calculate the asymptotic distribution of 
the time-averaged position $\ta{X}(t)$. Its ensemble mean falls onto the 
equilibrium mean, $\mean{\ta{X}(t)}\sim\meaneq{X}$. The bulk PDF of the 
fluctuations $\ta{\Delta X}(t)$ is given through 
Eq.~\eqref{taobs:pdf:bulk:scalinglimit},  where we insert the equilibrium 
quantities in Eqs.~\eqref{tax:mean} and~\eqref{tax:eqmom:constrained}. Plots are 
provided in Fig.~\ref{fig.tax.pdf}, left panel, for different values of the bias 
strength $\mu$. When the bias is stronger ($\mu$ larger), then the PDF of the 
fluctuations gets more peaked ($\kappa$ decreases) and more symmetric ($\beta>0$ 
decreases). 

The large fluctuations scaling limit is obtained by using Eqs.~\eqref{tax:mean} 
and~\eqref{tax:eqmom:constrained} in Eq.~\eqref{taobs:pdf:ldev:scalinglimit}. 
The respective plots can be found in the right panel of Fig.~\ref{fig.tax.pdf}. 
In contrast to the universal L\'evy-stable bulk statistics, the structure of the 
emergent infinite density does reveal two defining properties of the underlying 
CTRW. First, the reflective boundary condition causes the fluctuations to be 
bounded from the left, $-\meaneq{X}\leq\ta{\Delta X}(t)$. Second, the infinite 
density exhibits discontinuous steps at points where $\meaneq{X}+\ta{\Delta x}$ 
becomes an integer, an obvious signature of the discrete lattice structure. The 
effect is most pronounced for a strong bias, since then the CTRW samples  only a 
small number of lattice sites.

\section{Summary and outlook}\label{sec.outlook}
Within the framework of CTRW, we studied bounded, ergodic dynamics close to the 
non-ergodic phase. In this regime, equilibrium statistical mechanics, in 
particular the Boltzmann-Gibbs laws, dictate the values of  occupation times and 
time averages in the limit of infinite measurement times. But at finite times, 
the fluctuations of these quantities are significant and are to be analyzed on 
the footing of dual scaling laws, such as 
Eqs.~\eqref{otf:pdf:bulk:scalinglimit}-\eqref{otf:pdf:ldev:scalinglimit} 
or~\eqref{taobs:pdf:bulk:scalinglimit}-\eqref{taobs:pdf:ldev:scalinglimit}. Bulk 
fluctuations are to be distinguished from the rare, large fluctuations. The two 
emergent asymptotic scaling laws complement one another. Thus, the behavior of 
physical observables, like the $q$th order moments considered in 
Eqs.~\eqref{otf:moments}, are to be derived from bulk or large fluctuations, or 
both. 

The bulk fluctuations are asymptotically distributed according to asymmetric 
L\'evy-stable laws. This qualitative feature is universal for occupation times 
and general time averages alike, see Eq.~\eqref{otf:pdf:bulk:asymptotic} 
or~\eqref{taobs:pdf:bulk:asymptotic}. Only a single parameter, the skewness 
$\beta$, depends to some extent on the binding potential and the observable of 
interest. The rare large fluctuations also exhibit universality, in so far as 
they are generally described in terms of infinite densities, see 
Eq.~\eqref{otf:pdf:ldev:asymptotic} or~\eqref{taobs:pdf:ldev:asymptotic}. But 
the precise analytic form of the infinite density is much more sensitive to the 
details of the system and the particular observable under study. For instance, 
the natural bounds of the occupation times or time averages are only probed by 
the large fluctuations; also the discrete/continuous structure of the phase 
space is only reflected in the infinite density behavior.

Our equations connect the emergent L\'evy-stable laws and infinite densities 
with the occupation probability $\peq{x}$ of an equilibrium ensemble of 
independent random walkers. In a thermal system, we can thus relate the 
Boltzmann-Gibbs equilibrium statistics with the finite-time fluctuations around 
the ergodic limit. In the case of occupation times, this relationship is 
remarkably simple. For a given lattice site $x$, the single equilibrium 
occupation probability $\peq{x}$ controls the asymptotic distribution of both 
bulk and large fluctuations. Beyond that, only two dynamical parameters 
determine the distribution: the tail exponent $\alpha$ and the relative width 
$C=A/\mean{\tau}$ of the waiting time distribution~\eqref{broadtail}.

Hence, the equilibrium laws of statistical mechanics have a strong connection 
with asymmetric L\'evy-stable laws and infinite densities. We believe that 
similar relationships can be found in a vast variety of systems beyond CTRW. 
They derive from the  broad sojourn time distributions with power law tails as 
in Eq.~\eqref{broadtail}. This characteristic also appears in 
stochastic~\cite{Monthus1996,Burov2007} and 
deterministic~\cite{Zaslavsky2002,Korabel2009,Korabel2012,Akimoto2010,
Akimoto2012} models of complex motion, and has been measured 
experimentally~\cite{Scher1975,Weigel2011,Wong2004,Xu2011}. While the detailed 
asymptotic statistics of these systems might be complicated, our dual scaling 
approach may prove to be an effective analytical tool. Experimental studies of 
such systems may reveal the connection between model parameters ($\alpha$, $C$, 
$\peq{x}$) and other measurable quantities (temperature, particle size, 
confinement characteristics, etc.). An ambitious, yet desirable long-term goal 
is to embed the limit laws found here into large deviations 
theory~\cite{Touchette2009}. Furthermore, aging 
effects~\cite{Godreche2001,Schulz2013b} should be taken into consideration, if 
the CTRW-like process has been initiated long before we start to measure 
occupation times or time averages. We expect aging to be relevant in particular 
for the infinite density of large fluctuations. The regime $\alpha>2$ is an 
alternative direction to extend our studies. We believe that, in this case, an 
infinite density still provides the large fluctuation statistics, due to the 
divergence of moments $\mean{\tau^k}$ with $k>\alpha$. But how rare these large 
fluctuations are, and how their infinite density looks like or connects to the 
presumably Gaussian bulk distribution, are outstanding questions.

\acknowledgments
We acknowledge funding from the Israel Science Foundation.

\appendix

\section{Peaked contributions to the infinite density $\infdens{z}$ in 
Eq.~\eqref{otf:pdf:ldev:scalingfunction:differentiate}}
\label{appsec.otf:pdf:ldev}

The considerations that we expressed at the beginning of Sec.~\ref{sec.otf:ldev} 
have led us to exercise care when treating the outer boundaries $z=1$ in 
Eq.~\eqref{otf:pdf:ldev:scalingfunction:differentiate}, and respectively 
$\otf{x}{t}=-\peq{x}t,(1-\peq{x})t$ in Eq.~\eqref{otf:pdf:ldev:asymptotic}. 
Generally, derivatives of the step function $\indfu{z\leq1}$ lead to peaks of 
the form $\delta(z-1)$.  But for the CTRWs we studied in this work, such 
contributions are not present, as we show now. We abbreviate the expression 
$z^{-(\alpha-1)}$ appearing in 
Eq.~\eqref{otf:pdf:ldev:scalingfunction:differentiate} with $F(z)$ and write  
down the distributional identity
\begin{subequations}
\begin{align}
 \left\{\frac{\partial}{\partial z}\left[ \indfu{z\leq1} F(z)\right]\right\}
  &= \indfu{z\leq1} F'(z) -\delta(z-1) F(z) .
\end{align}
In principle, $F(z)$ could be any sufficiently well-behaved conventional 
function, which in particular does not have a pole at $z=1$. Similarly, we can 
write
\begin{align}
 &\left\{\frac{\partial^2}{\partial z^2}\left[ \indfu{z\leq1} 
F(z)\right]\right\}=
   \nn
  &\hspace{5mm}= \indfu{z\leq1} F''(z) -2\delta(z-1) F'(z)  +\delta'(z-1) F(z) 
\nn
  &\hspace{5mm}= \indfu{z\leq1} F''(z) -\delta(z-1) F'(z) +\delta(z-1) 
F(z)\frac{\partial}{\partial z} .
\end{align}
\label{app:differentiate}%
\end{subequations}
The derivative on the right-hand side acts upon any subjected test function. 
With respect to Eq.~\eqref{otf:pdf:ldev:scalingfunction:differentiate}, the 
above relation is useful for evaluating the expression containing 
$F(z)=(1-z)z^{-(\alpha-1)}$. 

With the identities~\eqref{app:differentiate}, we can continue 
Eq.~\eqref{otf:pdf:ldev:scalingfunction:differentiate} as
\begin{align}
 \infdens{z}
 &=\frac{\indfu{0<z\leq1}}{\alpha(\alpha-1)} \Big\{ -(\alpha-1)z^{-\alpha} + \nn
 &\hspace{15mm} 
+\left[(\alpha-1)\alpha+(2-\alpha)(\alpha-1)z\right]z^{-\alpha-1}\Big\} \nn
 &\quad -\frac{\delta(z-1)}{\alpha(\alpha-1)} \Big\{ 
z^{-(\alpha-1)}-\left[\alpha-1+(2-\alpha)z\right]z^{-\alpha}\Big\} \nn
 &\quad +\frac{\delta(z-1)}{\alpha(\alpha-1)} \Big\{ (1-z)z^{-(\alpha-1)} 
\Big\}\frac{\partial}{\partial z} \nn
 &=\indfu{0<z\leq1} z^{-1-\alpha} \left(1-\frac{\alpha-1}{\alpha}z\right) .
\end{align}
We observe that the $\delta$-peaks at the boundary $z=1$ indeed have zero 
contribution, and Eq.~\eqref{otf:pdf:ldev:asymptotic:scalingfct} holds true.

\section{Absolute moments of all orders, 
Eq.~\eqref{otf:moments}}\label{appsec.otf:moments}
We know that at large times $t$, the central part of the PDF $g(\otf{x}{t};t)$ 
is well approximated by $g_\subbulk(\otf{x}{t};t)$, while the outer parts are 
close to $g_\subldev(\otf{x}{t};t)$. Inside two intermediate overlap regions, 
both approximations feature the power law behavior given in 
Eq.~\eqref{otf:pdf:powerlaw}. We now introduce a positive, monotonic function 
$c(t)$, such that $c(t)$ and ${-c(t)}$ are bound to lie inside the positive and 
negative power law regions, respectively, i.e.
\begin{align}
  & {-t\peq{x}} \ll {-c(t)} \ll {-\kappa t^{1/\alpha}}  &\wedge&&
  \kappa t^{1/\alpha} \ll c(t) \ll t(1-\peq{x}) . 
\end{align}
Notice that the distance between the boundaries broadens as time increases. 
Therefore, we can always find a function $c(t)$ that remains inside this region, 
at least beyond some finite threshold time. We even have some freedom in 
defining its time dependence: $c(t)\simeq t^{1/\alpha}$ is fine, or $c(t)\simeq 
t$, or anything in between. We make use of this fact below.

\begin{widetext}
With the function $c(t)$ we can divide the value of an absolute moment into two 
contributions:
\begin{align}
 \mean{|\Otf{x}{t}|^q} 
 &= \left(\int_{|\otf{x}{t}|\leq c(t)} \right.+\left. \int_{|\otf{x}{t}|> 
c(t)}\right) |\otf{x}{t}|^q\, g(\otf{x}{t};t) \,d(\otf{x}{t}) \nn
 &\begin{aligned}
    &=\quad t^{q/\alpha} \int_{|y|\leq c(t)/t^{1/\alpha}} 
|y|^q\,t^{1/\alpha}g(yt^{1/\alpha};t)\,dy
    &+&\quad t^{-(\alpha-1)+q} \int_{|z|> c(t)/t} |z|^q\,t^\alpha g(zt;t)\,dz 
\nn
    &\sim\quad t^{q/\alpha} \int_{|y|\leq c(t)/t^{1/\alpha}} |y|^q\, 
g_\subbulk(y)\,dy
    &+&\quad t^{-(\alpha-1)+q} \int_{|z|> c(t)/t} |z|^q\, g_\subldev(z)\,dz
 \end{aligned}\\
 \label{moments:1}
\end{align}
On the last line, we used the asymptotic expressions as in 
Eqs.~{\eqref{otf:pdf:bulk:scalinglimit}} and~\eqref{otf:pdf:ldev:scalinglimit}. 
This is allowed, since the boundary $c(t)$ asserts that both integrals operate 
on regions where the respective approximation becomes more and more accurate 
with time. We now treat the three classes of moments separately.

\subsection{Low-order moments, $\boldsymbol{0<q<\alpha}$} 
We choose $c(t)\equiv t\,c_1$, with some $0<c_1\ll\min\{\peq{x},1-\peq{x}\}$, 
and continue Eq.~\eqref{moments:1}:
\begin{alignat}{2}
 \mean{|\Otf{x}{t}|^q} 
 &\sim\quad t^{q/\alpha} \int_{|y|\leq t^{1-1/\alpha}\,c_1} |y|^q\, 
g_\subbulk(y)\,dy
 &\quad&+\quad t^{-(\alpha-1)+q} \int_{|z|> c_1} |z|^q\, g_\subldev(z)\,dz \nn
 &=\quad t^{q/\alpha} \int_{-\infty}^\infty |y|^q\, g_\subbulk(y)\,dy
 &\quad&+\quad \Order{t^{-(\alpha-1)+q}} \nn
 &\sim\quad \mql \,t^{q/\alpha} \qquad(0<q<\alpha)
\end{alignat}
When going from the first to the second line, we used the fact that 
$g_\subbulk(y)$ is $\alpha$-stable, and therefore the related integral converges 
with time to the finite moment $\mql$ of order $q<\alpha$. In the third line, 
the bulk contribution dominates at large times, since we have the relations
\begin{equation}
 -(\alpha-1)+q \lessgtr q/\alpha \quad\xLeftrightarrow{\alpha>1}\quad q \lessgtr 
\alpha .
 \label{moments:dominant}
\end{equation}

\subsection{High-order moments, $\boldsymbol{q>\alpha}$.} We choose $c(t)\equiv 
t^{1/\alpha}\,c_2$, with some $\kappa\ll c_2$, and continue 
Eq.~\eqref{moments:1}:
\begin{alignat}{2}
 \mean{|\Otf{x}{t}|^q} 
 &\sim\quad t^{q/\alpha} \int_{|y|\leq c_2} |y|^q\, g_\subbulk(y)\,dy
 &\quad&+\quad t^{-(\alpha-1)+q} \int_{|z|> t^{1/\alpha-1}\,c_2} |z|^q\, 
g_\subldev(z)\,dz \nn
 &=\quad \Order{t^{q/\alpha}}
 &\quad&+\quad t^{-(\alpha-1)+q} \int_{-\peq{x}}^{1-\peq{x}} 
|z|^q\,g_\subldev(z;1) \,dz \nn
 &\sim\quad \mqg \,t^{-(\alpha-1)+q} \qquad(q>\alpha)
\end{alignat}
The integral on the second line gives the value of $\mqg$ as in 
Eq.~\eqref{otf:moments}. We stress that the latter is finite: the infinite 
density $g_\subldev(z)$ diverges at the origin like $|z|^{-\alpha-1}$, but the 
factor $|z|^q$ guarantees convergence for any $q>\alpha$. Finally, due to 
relation~\eqref{moments:dominant}, the large fluctuations with time dependence 
$t^{-(\alpha-1)+q}$ dominate higher order moments $q>\alpha$.

\subsection{Borderline case, $q=\alpha$} 
We set again $c(t)\equiv t^{1/\alpha}\,c_2$, $c_2\gg\kappa$. For $q=\alpha$, 
both bulk and large fluctuations contribute to moments in the same way. This can 
be seen as follows.
\begin{align}
 \mean{|\Otf{x}{t}|^\alpha} 
 &\sim\quad t \int_{|y|\leq c_2} |y|^\alpha\, g_\subbulk(y)\,dy
 \quad+\quad t \int_{|z|> t^{1/\alpha-1}\,c_2} |z|^\alpha\, g_\subldev(z)\,dz 
\nn
 &=\quad \Order{t}\quad -\quad t\,\frac{C\alpha}{|\Gamma(1-\alpha)|} \left[ 
(1-\peq{x})^\alpha\peq{x}\ln\left(t^{1/\alpha-1}\,c_2\right)  
+(1-\peq{x})\peq{x}^\alpha\ln\left(t^{1/\alpha-1}\,c_2\right)\right] \nn
 &\sim\quad 
\frac{C(\alpha-1)}{|\Gamma(1-\alpha)|}\left[(1-\peq{x})^\alpha\peq{x}+(1-\peq{x}
)\peq{x}^\alpha\right] \;t\ln t
\end{align}
\end{widetext}
Since $|z|^\alpha g_\subldev(z;1)$ behaves at the origin as $|z|^{-1}$, the 
large fluctuations integral on the first line diverges logarithmically as 
$t\rightarrow\infty$. This is why in this case, we had to consider explicitly 
the form of the power law pole, Eq.~\eqref{otf:pdf:powerlaw}. Terms of the order 
of $t$, were neglected on the last line. This includes in particular the 
contributions from the very center of the stable density ($|y|\leq c_2$) and the 
extremities of the infinite density ($z\approx -\peq{x}$ and 
$z\approx1-\peq{x}$) and the $c_2$-dependence of function $c(t)$. Hence, in a 
sense, only the power law overlap between the L\'evy-stable and the infinite 
density determines the asymptotic form of the moment $q=\alpha$.

\section{Fluctuations statistics of general time averages, 
Eqs.~\eqref{taobs:pdf:bulk:scalinglimit}-\eqref{taobs:pdf:ldev:asymptotic}}
\label{appsec.taobs}
We outline here the derivation of the long-time asymptotics of the PDF 
$h(\ta{\obsf};t)$ for the fluctuations $\ta{\Obsf}(t)$ of the time average of a 
general observable $\Obs$. Our starting point is the joint PDF 
$f(\ta{\obs},n;t)$ for measuring the $\ta{\Obs}(t)=\ta{\obs}$ at time $t$, while 
having made exactly $N(t)=n$ random walk steps up to that point. We can write 
down its Fourier-Laplace representation as
\begin{align}
 \trafo{f}(u,n;s) &=\int_0^\infty e^{-st} 
\mean{\expO{-ut\ta{\Obs}(t)}\delta[N(t)=n]}\,dt .
\end{align}
The average $\mean{\cdot}$ is to be interpreted here as an average over all 
realizations of the CTRW process. According to Ref.~\cite{Rebenshtok2008}, 
Sec.~5.3, one can write
\begin{align}
 &\trafo{f}(u,n;s) = \nn&\;=
 \Bigg\langle 
\frac{1-\trafo{\psi}\left(s+u\obs_{X^\suprw(n)}\right)}{s+u\obs_{X^\suprw(n)}} 
 \prod_x \left[\trafo{\psi}(s+u\obs_x)\right]^{N^\suprw_x(n)} \Bigg\rangle
 . 
 \label{app:taobs:pdfn}
\end{align}
This should be read as follows. The average with respect to the random, 
independent and identically distributed waiting times has already been carried 
out. The statistics of waiting times is embedded via the Laplace transform 
$\trafo{\psi}(s)$. The remaining average is to be taken with respect to two 
aspects of the trajectories of an analogue \emph{ordinary} random walk (RW), 
i.e. where the time between steps is fixed to unity. First, we have to average 
over all terminal positions $X^\suprw(n)$ after $n$ steps. Second, we average 
over the possible sets of visitation numbers $N^\suprw_x(n)$, i.e. the number of 
times an ordinary random walker has visits each of the sites $x$ of the lattice 
within $n$ steps. 

We want to approximate Eq.~\eqref{app:taobs:pdfn} in the limit of large times 
and, consequently, large values of $t\ta{\Obs}(t)$. This translates to small 
$s$, $u$. On the one hand, for the broad-tailed waiting times as in 
Eq.~\eqref{broadtail}, this means we expand
\begin{align}
 \trafo{\psi}(s+u\Obs)&\sim 1-\mean{\tau}(s+u\Obs)+A(s+u\Obs)^\alpha 
 .\label{app:taobs:psiL}
\end{align}
On the other hand, it is natural to assume that we need to concentrate on large 
$n$, so we require the statistics of the ordinary random walk after many steps. 
Since we confine our discussion to ergodic random walks, it is easy to say what 
happens when $n$ becomes infinite. The position $X^\suprw(n)$ is distributed 
according to the $n$-independent equilibrium distribution $\peq{x}$, while the 
visitation numbers converge as $N^\suprw(n)/n\rightarrow\peq{x}$. At finite $n$, 
deviations from these ergodic limits appear. However, for ordinary random walks, 
these kind of fluctuations do typically not exhibit broad-tailed statistics. We 
rather assume that their contribution in Eq.~\eqref{app:taobs:pdfn} is of the 
order $\Order{(s+u\Obs)^2}$. They are then negligible as compared to the 
fluctuations caused by the broad-tailed waiting times~\eqref{app:taobs:psiL}. 
Recall that in the specific case of the site occupation times $\Otf{x}{t}$, 
Sec.~\ref{sec.otf}, we gave a more rigorous argument for this assumption in 
terms of a first-passage time analysis. Only the \emph{mean} of the 
first-passage time of the ordinary random walk plays a role in the asymptotics 
of occupation time fluctuations. For this or similar arguments to hold, it is 
obviously crucial that $\alpha<2$.

We thus now approximate Eq.~\eqref{app:taobs:pdfn} by using 
Eq.~\eqref{app:taobs:psiL}, by writing the average with respect to $X^\suprw(n)$ 
as an equilibrium average and by replacing the visitation numbers $N^\suprw(n)$ 
by their ergodic limit values $n\peq{x}$. This yields
\begin{align}
 &\trafo{f}(u,n;s) \sim \nn
 &\quad\sim  
\meaneq{\frac{\mean{\tau}\left(s+u\Obs\right)-A\left(s+u\Obs\right)^\alpha}{
s+u\Obs}}
 \times \nn 
 &\quad\quad
 \prod_x \left[ 
1-\mean{\tau}\left(s+u\obs_x\right)+A\left(s+u\obs_x\right)^\alpha 
\right]^{n\peq{x}} \nn
 &\quad= 
 \left[ \mean{\tau}-A\meaneq{(s+u\Obs)^{\alpha-1}} \right] \times \nn
 &\quad\quad
 \exp\!\left\{ 
n\sum_x\peq{x}\ln\left[1-\mean{\tau}(s+u\obs_x)+A(s+u\obs_x)^\alpha\right] 
\right\}
 . 
\end{align}
At this point, we marginalize out the number of steps. Since we are operating in 
a limit where $n$ is assumed large, we find it appropriate to treat $n$ as 
continuous and integrate, that is
\begin{align}
 \trafo{f}(u;s)&\sim\int_0^\infty\trafo{f}(u,n;s)\,dn = \nn 
&=\frac{\mean{\tau}-A\meaneq{(s+u\Obs)^{\alpha-1}}}{\sum_x\peq{x}\ln\left[
1-\mean{\tau}(s+u\obs_x)+A(s+u\obs_x)^\alpha\right]} \nn
 &\sim 
\frac{1-C\meaneq{(s+u\Obs)^{\alpha-1}}}{s+u\meaneq{\Obs}-C\meaneq{(s+u\Obs)^{
\alpha}}} .
 \label{app:taobs:pdf:completeL}
\end{align}
In the last step, we used $\ln(1+x)\sim x+\Order{x^2}$ for small $x$. 
Equation~\eqref{app:taobs:pdf:completeL} is the generalization of 
Eq.~\eqref{ot:pdf:complete:trafo}. The latter can be recovered by considering 
the occupation observable in Eq.~\eqref{taobs:otf}.

The PDF of the fluctuations $\ta{\Obsf}(t)$ are most suitably discussed in 
Fourier-Laplace space. By shifting by $-t\meaneq{\Obs}$, we get
\begin{align}
 \trafo{h}(k;t) &= \int_0^\infty\mean{\expO{ikt\ta{\Obsf}(t)}}\,dt= 
\trafo{f}(-ik;s+ik\meaneq{\Obs}) \nn
 &\sim 
\frac{1-C\meaneq{(s-ik\Obsf)^{\alpha-1}}}{s-C\meaneq{(s-ik\Obsf)^{\alpha}}} ,
 \label{app:taobsf:pdf:completeL}
\end{align}
generalizing Eq.~\eqref{otf:pdf:complete:trafo}. By methods which are completely 
analogous to those used in Sec.~\ref{sec.otf:bulk} one can compute the limit 
where $s$ becomes small while $s/|k|^\alpha$ is fixed. This gives the L\'evy 
$\alpha$-stable bulk statistics of the fluctuations of time averages, 
Eqs.~\eqref{taobs:pdf:bulk:asymptotic}-\eqref{taobs:eqmom}. The limit where $s$ 
is small and $s/|k|$ fixed is obtained through the  procedure discussed in 
Sec.~\ref{sec.otf:ldev} and~\ref{appsec.otf:pdf:ldev}. It yields the infinite 
density of large fluctuations, Eq.~\eqref{taobs:pdf:ldev:asymptotic}.


\begin{thebibliography}{99}

\bibitem{Agmon2010}
N. Agmon, Chem. Phys. Lett. \textbf{497}, 184 (2010).

\bibitem{Berezhkovskii1998}
A. M. Berezhkovskii, V. Zaloj, and N. Agmon, Phys. Rev. E \textbf{57}, 3937
(1998).

\bibitem{Benichou2003}
O. B\'enichou, M. Coppey, J. Klafter, M. Moreau, and G. Oshanin, J. Phys. A 
\textbf{36} 7225 (2003).

\bibitem{Godreche2001} C. Godr{\`e}che and J. M. Luck, J. Stat. Phys. 
\textbf{104}, 489 (2001).

\bibitem{Majumdar1999}
S. N. Majumdar, Curr. Sci. \textbf{77}, 370 (1999).

\bibitem{Bardou2002} F. Bardou, J.-P. Bouchaud, A. Aspect, and C. 
Cohen-Tannoudji,
\emph{L\'evy Statistics and Laser Cooling\/} (Cambridge University Press,
Cambridge, UK, 2002).

\bibitem{Bardou1994} F. Bardou, J.-P. Bouchaud, O. Emile, A. Aspect, and C.
Cohen-Tannoudji, Phys. Rev. Lett. \textbf{72}, 203 (1994).

\bibitem{Jarzynski2011}
C. Jarzynski, Annu. Rev. Condens. Matter Phys. \textbf{2}, 329 (2011)
Thermodynamics at the Nanoscale

\bibitem{Seifert2012}
U. Seifert, Rep. Prog. Phys. \textbf{75}, 126001 (2012).

\bibitem{Monthus1996} C. Monthus, and J.-P. Bouchaud, J. Phys. A \textbf{29}, 
3847 (1996).

\bibitem{Bel2005}
G. Bel, and E. Barkai, Phys. Rev. Lett. \textbf{94}, 240602 (2005).

\bibitem{Bel2006}
G. Bel, and E. Barkai, Phys. Rev. E \textbf{73}, 016125 (2006)

\bibitem{Majumdar2002}
S. N. Majumdar, and A. Comtet, Phys. Rev. Lett. \textbf{89}, 060601 (2002).

\bibitem{Burov2007}
S. Burov, and E. Barkai, Phys. Rev. Lett. \textbf{98}, 250601 (2007)

\bibitem{Scher1975} 
H. Scher and E. W. Montroll, Phys. Rev. B \textbf{12}, 2455 (1975).

\bibitem{Lutz2004} E. Lutz, Phys. Rev. Lett. \textbf{93}, 190602 (2004).

\bibitem{Jeon2011} 
J.-H. Jeon,  V. Tejedor, S. Burov, E. Barkai, C. Selhuber-Unkel, K. 
Berg-S{\o}rensen, L. Oddershede, and R. Metzler, Phys. Rev. Lett. \textbf{106}, 
048103 (2011).

\bibitem{Weigel2011}
A. V. Weigel, B. Simon, M. M. Tamkun, and D. Krapf, PNAS \textbf{108},
6438 (2011).
single-molecule tracking

\bibitem{Xu2011} Q. Xu, L. Feng, R. Sha, N. C. Seeman, and P. M. Chaikin, Phys. 
Rev. Lett. \textbf{106}, 228102 (2011).

\bibitem{Wong2004}
I. Y. Wong, M. L. Gardel, D. R. Reichman, E. R. Weeks, M. T. Valentine, A. R. 
Bausch, and D. A. Weitz, Phys. Rev. Lett. \textbf{92}, 178101 (2004).
Networks

\bibitem{Bouchaud1990}
J.-P. Bouchaud, A. Georges, Phys. Rep. \textbf{195}, 127 (1990) 
PHYSICAL APPLICATIONS

\bibitem{Metzler2000} 
R. Metzler and J. Klafter, Phys. Rep. \textbf{339}, 1 (2000).

\bibitem{Metzler2004}
R. Metzler and J. Klafter, J. Phys. A \textbf{37}, R161 (2004).
description of anomalous transport by fractional dynamics

\bibitem{Metzler2014}
R. Metzler, J.-H. Jeon, A. G. Cherstvy, and E. Barkai, Phys. Chem. Chem. Phys. 
\textbf{16}, 24128 (2014)
non-ergodicity, and ageing at the centenary of single particle tracking

\bibitem{RaWRaE} 
B. D. Hughes, \textit{Random Walks and Random Environments, Vol. I: Random 
Walks} (Oxford University Press, Oxford, UK, 1995).

\bibitem{Bertin2008}
E. Bertin, and F. Bardou, Am. J. Phys. \textbf{76}, 630 (2008).

\bibitem{Thaler1983}
M. Thaler, Israel J. Math. \textbf{46}, 67 (1983).
points

\bibitem{Thaler2006}
M. Thaler and R. Zweim\"uller, Probability theory and related fields 
\textbf{135}, 15 (2006).

\bibitem{InfErgTheo}
J. Aaronson, \textit{An Introduction to Infinite Ergodic Theory} (American 
Mathematical Society, Providence, 1997).

\bibitem{Korabel2009}
N. Korabel, and E. Barkai, Phys. Rev. Lett. \textbf{102}, 050601 (2009).

\bibitem{Korabel2012}
N. Korabel, and E. Barkai, Phys. Rev. Lett. \textbf{108}, 060604 (2012).
Chaos

\bibitem{Akimoto2010}
T. Akimoto, and T. Miyaguchi, Phys. Rev. E. \textbf{82}, 030102(R) (2010).

\bibitem{Akimoto2012}
T. Akimoto, Phys. Rev. Lett. \textbf{108}, 164101 (2012).

\bibitem{Kessler2010}
D. A. Kessler, and E. Barkai, Phys. Rev. Lett. \textbf{105}, 120602 (2010).
Lattices

\bibitem{Holz2013}
P. C. Holz, A. Dechant, and E. Lutz, Europhys. Lett. \textbf{109}, 23001 (2015).

\bibitem{Lutz2013}
E. Lutz and F. Renzoni, Nature Phys. \textbf{9}, 615 (2013).

\bibitem{Rebenshtok2014}
A. Rebenshtok, S. Denisov, P. H\"anggi, and E. Barkai, Phys. Rev. Lett. 
\textbf{112}, 110601 (2014), 
Limit Theorem
\textit{ibid.} Phys. Rev. E \textbf{90}, 062135 (2014).

\bibitem{Schulz2013b}
J. H. P. Schulz, E. Barkai, and R. Metzler, Phys. Rev. X \textbf{4}, 011028 
(2014).

\bibitem{Rebenshtok2008}
A. Rebenshtok, and E. Barkai, J. Stat. Phys. \textbf{133}, 565 (2008).

\bibitem{Burov2010a}
S. Burov, R. Metzler, and E. Barkai, PNAS \textbf{107}, 13228 (2010).

\bibitem{Kac}
M. Kac, \textit{Probability and Related Topics in Physical Sciences} 
(Interscience, New York, 1958).

\bibitem{Zaslavsky2002}
G. M. Zaslavsky, Phys. Rep. \textbf{371}, 461 (2002).

\bibitem{GCLT}
B. V. Gnedenko and A. N. Kolmogorov, \textit{Limit Distributions for Sums of 
Independent Random Variables} (Addison-Wesley, Cambridge, MA, 1954).

\bibitem{footnote:initcond}
We note that Eq.~\eqref{ot} is derived under the assumption that at time $t=0$, 
the random walker is not residing on $x$. However, in the large time limit, the 
choice of this initial condition has no effect (see Ref.~\cite{Bel2006}).

\bibitem{Benichou2014}
O. B\'enichou, and R. Voituriez, Physics Reports \textbf{539}, 225 (2014).
kinetics

\bibitem{footnote:mfpt}
The authors of Ref.~\cite{Benichou2014} use a definition of the first-return 
time $\bar{\tau}_\subnx^\suprw$ which slightly differs from ours: it includes 
the time spent at the site $x$ before jumping off. Hence, for us, 
$\tau_\subnx^\suprw=\bar{\tau}_\subnx^\suprw-1$.

\bibitem{StableProc}
G. Samorodnitsky and M. S. Taqqu, \textit{Stable Non-Gaussian Random Processes: 
Stochastic Models with Infinite Variance} (Chapman \& Hall, 1994).

\bibitem{Vulpiani1999}
P. Castiglione, A. Mazzino, P. Muratore-Ginanneschi, and
A. Vulpiani, Physica D (Amsterdam) \textbf{134}, 75 (1999).

\bibitem{Touchette2009}
H. Touchette, Phys. Rep. \textbf{478}, 1 (2009).

\end{thebibliography}
\end{document}